\documentclass[prd,preprintnumbers,amsmath,amssymb,preprint,nofootinbib,eqsecnum]{revtex4}
\usepackage{latexsym,ifthen,graphics,color,hyperref,epsfig}

\newcommand{\nothing}{}
\newlength{\eqnheight}

\newcommand{\ie}{{i.e.}}
\newcommand{\cf}{{cf.}}
\newcommand{\eg}{{e.g.}}
\newcommand{\aka}{{a.k.a.}}
\newcommand{\etc}{{etc.}}
\newcommand{\viz}{{viz.}}
\newcommand{\wrt}{with respect to}

\newcommand{\lhs}{left-hand side}

\newcommand{\rhs}{right-hand side}

\newcommand{\Kahler}{K\"{a}hler}

\newcommand{\be}{\begin{equation}}
\newcommand{\ee}{\end{equation}}
\newcommand{\bea}{\begin{eqnarray}}
\newcommand{\eea}{\end{eqnarray}}
\newcommand{\beas}{\begin{eqnarray*}}
\newcommand{\eeas}{\end{eqnarray*}}

\newcommand{\bear}{\begin{array}{l}}
\newcommand{\eear}{\end{array}}

\newcommand{\bcf}{\begin{center}\begin{figure}}
\newcommand{\ecf}{\end{figure}\end{center}}

\newcommand{\bct}{\begin{center}\begin{table}}
\newcommand{\ect}{\end{table}\end{center}}

\newcommand{\ds}{\displaystyle}

\def\eq#1{(\ref{eq:#1})}
\def\eqn#1{equation~(\ref{eq:#1})}

\def\eqs#1#2{(\ref{eq:#1}) and~(\ref{eq:#2})}

\def\eqns#1#2{equations~(\ref{eq:#1}) and~(\ref{eq:#2})}

\def\sec#1{section~\ref{sec:#1}}

\def\secs#1#2{sections~\ref{sec:#1} and~\ref{sec:#2}}
\def\appx#1{appendix~\ref{app:#1}}
\def\fig#1{figure~\ref{fig:#1}}

\def\app#1{appendix~\ref{app:#1}}

\newcommand{\Int}[1]{\int \!\! d^D \! #1 \,}
\newcommand{\FourInt}[1]{\int \!\! d^4 \! #1 \,}
\newcommand{\TwoInt}[1]{\int \!\! d^2  #1 \,}
\newcommand{\MomInt}[2]{\int \!\! \frac{d^{#1} #2}{(2\pi)^{#1}} \, }

\newcommand{\FermInt}[2]{\int \!\! d^{#1} #2 \, }

\newcommand{\FourVol}[1]{d^4 \! #1 \,}

\newcommand{\der}[2]{\ensuremath{\frac{d #1}{d #2}}}

\newcommand{\pder}[2]{\ensuremath{\frac{\partial #1}{\partial #2}}}

\newcommand{\fder}[2]{\ensuremath{\frac{\delta #1}{\delta #2}}}

\newcommand{\Or}{\mathcal{O}}

\newcommand{\order}[1]{\Or ( #1 )}
\newcommand{\hf}{\frac{1}{2}}

\newcommand{\SU}{\mathrm{SU}}
\newcommand{\SO}{\mathrm{SO}}

\newcommand{\firstspinor}[1]{#1;}
\newcommand{\secondspinor}[1]{;#1}

\def\dd{\dot{\Delta}}

\def\hS{\hat{S}}

\def\one{\hbox{1\kern-.8mm l}}

\newcommand{\anti}[1]{\overline{#1}}
\newcommand{\fermprod}[2]{(#1 #2)}

\newcommand{\suf}{\Phi}
\newcommand{\asuf}{\anti{\Phi}}
\newcommand{\ptlsf}{\phi}
\newcommand{\ptlasf}{\anti{\phi}}
\newcommand{\SR}{S^{\mathrm{I}}}
\newcommand{\hSR}{\hS^{\mathrm{I}}}
\newcommand{\SigmaR}{\Sigma^{\mathrm{I}}}

\newcommand{\alphadot}{\dot{\alpha}}

\newcommand{\Dbar}{\anti{D}}
\newcommand{\thetabar}{\anti{\theta}}
\newcommand{\sigmabar}{\anti{\sigma}}
\newcommand{\rhobar}{\anti{\rho}}
\newcommand{\kappabar}{\anti{\kappa}}
\newcommand{\omegabar}{\anti{\omega}}
\newcommand{\zetabar}{\anti{\zeta}}
\newcommand{\ybar}{\anti{y}}
\newcommand{\Kah}{K}
\newcommand{\SP}{f}
\newcommand{\SPProj}{\mathcal{P}_{\SP}}
\newcommand{\KPProj}{\mathcal{P}_0}

\newcommand{\dual}{\mathcal{D}}
\newcommand{\dualv}[1]{\dual^{(#1)}}
\newcommand{\dualvm}[1]{\dual_m^{(#1)}}

\newcommand{\dopi}{\Irr}
\newcommand{\dopiv}[1]{\Irr^{(#1)}}

\newcommand{\dualvpr}[1]{\dual'^{(#1)}}

\newcommand{\ztilde}{\tilde{z}}

\newcommand{\Irr}{\mathcal{I}}

\newcommand{\itp}{\Delta^{-1}}

\newcommand{\DiagDot}{\scriptstyle \bullet}
\newcommand{\DummyKernel}{\ensuremath{\stackrel{\bullet}{\mbox{\rule{1cm}{.2mm}}}}}

\newcommand{\flow}{\Lambda \partial_\Lambda}

\newcommand{\totalflow}{\Lambda \der{}{\Lambda}}

\newcommand{\dec}[3][0]{\ensuremath{\left[ #2 \hspace{#1em} \right]^{#3}}}

\newcommand{\dep}{\tilde{\Delta}}

\newlength{\Strutheight}

\newcounter{Diagrams}
\Alph{Diagrams}
\newtheorem{Diag}{}[Diagrams]

\newlength{\VertexWidth}

\newlength{\LabLength}

\newcommand{\arXiv}[2]{arXiv:{#1} [#2]}

\newcommand{\hepth}[1]{hep-th/#1}
\newcommand{\hepph}[1]{hep-ph/#1}
\newcommand{\heplat}[1]{hep-lat/#1}

\begin{document}

\preprint{DIAS-STP-08-11}

\title{On the Renormalization of Theories of a Scalar Chiral Superfield}

\author{Oliver J.~Rosten}
\email{orosten@stp.dias.ie}
\affiliation{Dublin Institute for Advanced Studies, 10 Burlington Road, Dublin 4, Ireland}

\begin{abstract}

	An exact renormalization group 
	for theories of a scalar chiral superfield
	is formulated, directly in four dimensional Euclidean space.
	By constructing a projector which isolates the
	superpotential from the full Wilsonian effective action, it is shown that the nonperturbative 
	nonrenormalization theorem follows, quite simply, from the flow equation.
	Next, it is argued that there do not exist any physically acceptable non-trivial fixed points.	
	Finally, the Wess-Zumino model is considered, as a low energy effective theory.
	Following an evaluation of the one and two loop $\beta$-function
	coefficients, to illustrate the ease of use of the formalism, it is shown that the $\beta$-function
	in the massless case does not receive any nonperturbative power corrections. 
\end{abstract}


\maketitle

\tableofcontents

\section{Introduction}

A crucial question that should be asked of any quantum field theory is whether or not
it is renormalizable. However, to definitively answer this question is often 
far from easy. A case in point is scalar field theory in $d=4$ dimensions. Let us start by supposing
that we introduce an overall momentum cutoff, $\Lambda_0$, the `bare scale'. Now, without any
further restrictions, there are an infinite number of different theories we could
consider, corresponding to different choices of the bare interactions. At least within perturbation
theory, one
such choice appears to be special: if we take just a mass term and a
$\lambda \varphi^4$ interaction then it is very well known that the theory is perturbatively
renormalizable. In other words, if we send $\Lambda_0 \rightarrow \infty$ (\aka\ taking the continuum
limit), then all
ultraviolet (UV) divergences can be absorbed into just the two couplings and the anomalous dimension of the field. However, beyond
perturbation theory, this breaks down. For example, defining this $\lambda \varphi^4$ 
theory on a lattice,
it can be (essentially) proven that the only continuum limits are trivial~\cite{Frohlich-Trivial}. 

The resolution to this apparent paradox is that taking the limit $\Lambda_0 \rightarrow \infty$
within perturbation theory amounts to a sleight of hand. Imagine integrating out degrees of
freedom between the bare scale and a much lower, effective scale, $\Lambda$. 
The point is that perturbation theory done at the scale $\Lambda$ is in fact only correct
up to $\order{\Lambda/\Lambda_0}$ terms. Formally, one can send $\Lambda_0 \rightarrow \infty$,
after which all quantities can be written in `self-similar' form~\cite{TRM-Elements,B+B}: \ie\ the results of all
perturbative calculations can be expressed as functions of the renormalized couplings, $m(\Lambda)$ and $\lambda(\Lambda)$, and the anomalous dimension, $\eta(\Lambda)$.
Indeed, self-similarity is precisely a statement of
renormalizability, since nothing has any explicit dependence on $\Lambda/\Lambda_0$.
The sleight of hand has come about because the various perturbative series are not, by
themselves, well defined: when one attempts to resum the
hopefully asymptotic perturbative series using \eg\ the Borel transform, it is found that there
are poles on the positive real axis of the Borel plane, impeding this 
procedure.\footnote{
Poles of this type can have different origins; those arising due to small/large loop momentum behaviour
are known as renormalons---for a review see~\cite{BenekeReview}. For theories which are
perturbatively renormalizable but for which an interacting continuum
limit based around the Gaussian fixed point nevertheless does not exist, ultraviolet (UV) renormalons give rise to the poles along the positive real axis.}

Whilst one
can avoid these poles by deforming the contour of integration, there is an ambiguity relating to
whether the contour goes above, or below, each pole. To arrive at an \emph{unambiguous} result one must include the $\Lambda/\Lambda_0$ terms which
were earlier thrown away. Doing so manifestly spoils self-similarity and hence renormalizability.

If we define the $\beta$-function, as usual, according to
\be
	\beta \equiv \Lambda \der{\lambda}{\Lambda},
\label{eq:betadefn}
\ee
and denote the one-loop $\beta$-function by $\beta_1$ then it is apparent that the 
$\Lambda/\Lambda_0$
contributions are indeed nonperturbative:
\be
	\Lambda/\Lambda_0 \sim e^{-1/2 \beta_1 \lambda^2}.
\label{eq:PowerCorr}
\ee
So, it is quite possible that perturbative conclusions about renormalizability differ from the nonpertubative ones.  Consequently, it is quite consistent that the perturbatively renormalizable
$\lambda \varphi^4$ model does not strictly have an interacting continuum limit. But what about 
all the other possible models we could have written down at the bare scale?

At first sight, answering this question is nigh impossible: after all, we can hardly check
every single such model to see whether, nonperturbatively, an interacting continuum limit exists. Fortunately, the question can be rephrased in a different way which, whilst still hard to answer in general, is
nevertheless 
much more amenable to solution. To do this, we must adopt Wilson's picture of
renormalization, whereby nonperturbatively renormalizable theories follow directly
from critical fixed points of the renormalization group (RG) and the `renormalized trajectories'
emanating from them~\cite{Wilson}. The first point to make is that critical fixed points correspond
to conformal field theories. These theories are therefore renormalizable in the nonperturbative sense: since they are scale independent, they must be independent of $\Lambda_0$, which can thus
be trivially sent to infinity. 

It is very simple to show, nonperturbatively, that scale dependent renormalizable theories follow by considering flows out of some critical fixed point along the relevant and marginally relevant\footnote{Henceforth, we will take `relevant' to include marginally relevant.}
directions \emph{as defined at this fixed point}~\cite{TRM-Elements}. A crucial feature of
these renormalized trajectories is that they are strictly self-similar, and this is a direct reflection of
nonperturbative renormalizability. 

Thus, rather than considering all possible theories at the bare scale and seeing whether
a continuum limit exists, we instead search for critical fixed points. If we find only
the Gaussian one, then we know that no interacting continuum limits can exist in
$d=4$ scalar field theory: with respect to this fixed point, the only relevant direction is the mass; $\lambda$ is marginally
\emph{irrelevant} and all other directions are even more irrelevant still. However, if a
non-trivial fixed point is found, then everything changes. If this fixed point were to have
relevant directions, then these could be used to construct a continuum limit. Now, suppose
that there exist  RG flows from this putative fixed which take us down towards the Gaussian fixed point. As
we begin our journey into the infrared, at some point we pass the scale we earlier denoted by $\Lambda_0$.
We can, if we choose, still call the action at this scale the bare action. But now it is
\emph{determined} by our choice of renormalized trajectory (this information is encoded in
the integration constants associated with the relevant directions). It is for this reason
that the bare action along a renormalized trajectory is sometimes referred to as the `perfect action' in the
vicinity of the UV fixed point~\cite{Perfect}. Continuing our journey, we ultimately reach the vicinity
of the Gaussian fixed point. Here, all interactions die away, with the exception of the
mass, which is relevant at the Gaussian fixed point. However, of the other interactions,
$\lambda$ dies away by far the slowest (logarithmic decay, compared to power law decay)
and so, sufficiently close to the Gaussian fixed point, we are effectively back to 
a $\lambda \varphi^4$ model. Indeed, this model is the good low energy effective
theory; but note that, crucially, all other interactions would have to be retained if
one wished to reconstruct the RG trajectory back into the UV.

This scenario, whereby a low energy effective theory is the result of a flow down from
a UV fixed point is often called asymptotic safety~\cite{Weinberg-AS}. Recently,
however, such a scenario was ruled out for scalar field theory in $d \geq 4$
as it was shown that no physically acceptable non-trivial fixed points
exist~\cite{Trivial}.  
There are two criteria that were used---and which we shall use in this paper---to 
determine the physical acceptability of a fixed point. The
first is `quasi-locality'~\cite{ym}: we demand that the action has an all orders derivative
expansion. Given that the analysis of~\cite{Trivial} was performed in 
Euclidean space, the second is 
that the theory makes sense as a unitary quantum field theory, upon continuation to Minkowski space.

The analysis of critical 
fixed points in scalar field theory presented in~\cite{Trivial} proceeded in two steps, depending on
the sign of the anomalous dimension, $\eta_\star$ (we will use $\star$ to denote fixed point quantities). First, fixed points
with $\eta_\star \geq 0$ were considered. (For $d=4$, in the case
where $\eta_\star =0$, Pohlmeyer's theorem~\cite{Pohlmeyer} implies that the only
critical fixed point is the Gaussian one.) For $\eta_\star \geq 0$, it was demonstrated
that no non-trivial fixed points exist in $d \geq 4$. As for fixed points with
$\eta_\star < 0$, it was shown that, should such fixed points
exist, then they are necessarily non-unitary since the 
kinetic term lacks the standard $p^2$ part. This can be seen explicitly for
the exotic Gaussian fixed points discovered by Wegner~\cite{Wegner_CS}.

The aim of this paper is to explore various aspects of the renormalizability of 
theories of a scalar chiral superfield in four dimensions. In line with the previous discussion, we
avoided explicit mention of the Wess-Zumino model in the previous sentence. 
As before, this is because in this supersymmetric case
\begin{enumerate}
	\item it is very well known that the Gaussian fixed point does not support
	interacting renormalized trajectories;

	\item there are no interacting continuum limits of the Wess-Zumino model.
\end{enumerate}
The latter fact can be deduced much more straightforwardly~\cite{Zumino-Trivial,Nappi} 
than in the case of $d=4$ scalar field theory, on account of the nonrenormalization theorem~\cite{nrts}
and Pohlmeyer's theorem. Indeed, we can state in complete generality that there
cannot be \emph{any} non-trivial fixed point with a three-point superpotential coupling, $\lambda$, as we now discuss.
(Henceforth, we exclusively use $\lambda$ to denote this coupling.)

The first point to make is that, to uncover fixed point behaviour, we should rescale to dimensionless
variables by dividing all quantities by $\Lambda$ raised to the appropriate scaling dimension.
This means that the superpotential does now renormalize, but only via the scaling dimension
of the field. In particular, the three-point superpotential coupling, which has zero canonical dimension,
acquires a scaling from the anomalous dimension of the field. Now, at a fixed point, all
couplings must stop flowing, by definition. Therefore, if the fixed point action possesses a three-point superpotential term, the anomalous dimension must vanish. But Pohlmeyer's theorem implies that any critical fixed point (in integer dimension) with vanishing anomalous dimension must be the trivial one.

Of course, this says nothing as to the existence, or otherwise, of  non-trivial fixed points without a three-point superpotential term. Moreover, such fixed points could potentially furnish an asymptotic safety
scenario for the Wess-Zumino model: 
since we are working in dimensionless variables, $\lambda$ does scale and
so can in principle be a relevant direction at a fixed point (this is no different from saying that the mass
is relevant at the Gaussian fixed point, despite the fact that there are no quantum corrections along the trivial mass direction). However, if such fixed points are to exist, it was recently shown that
they can only be used to construct an asymptotic safety scenario for the Wess-Zumino model if
the fixed point has
\begin{enumerate}
	\item negative anomalous dimension;
	
	\item at least one relevant direction coming from the \Kahler\ potential.
\end{enumerate}
The proof of this is very simple, utilizing only the nonrenormalization theorem and
Pohlmeyer's theorem~\cite{Safety}.

However, by adapting the methodology of~\cite{Trivial}, we will show that, should
any fixed points with negative anomalous dimension exist, they necessarily
correspond to non-unitary theories. Consequently, an asymptotic safety scenario
for the Wess-Zumino model is ruled out. Furthermore, it will be shown that there
are no physically acceptable non-trivial fixed points with positive anomalous
dimension, either (just because such a fixed point cannot possess a trajectory that flows towards the Wess-Zumino action does not mean that such a fixed point cannot exist; a separate argument is required to show this). Thus, an asymptotic safety scenario is ruled out for general
theories of a scalar chiral superfield.

In addition to this comprehensive study of the non-existence of useful fixed points,
a new proof of the nonperturbative renormalization theorem will be provided. 
It is not as elegant as Seiberg's beautiful argument~\cite{nrts} but it has the advantage
of being less heuristic, as it follows directly (and, it should be added, rather simply) from
the flow equation. 

Finally, the $\beta$-function of the Wess-Zumino model---considered as a low energy effective 
theory---is studied.
First, an explicit computation of the one and two-loop coefficients is provided, to illustrate the
ease of use our approach which, we note, is formulated directly in $d=4$. Secondly, we
adapt an analysis performed in QED~\cite{Resum} to show that the $\beta$-function in the massless
model (given the definition of the coupling implicit in the approach) is free of nonperturbative power corrections and hence is expected to be (Borel) 
resummable.

The formalism that will be employed throughout this paper is the Exact Renormalization Group (ERG), which is essentially the continuous version of Wilson's RG~\cite{Wilson,WH}.
Central to the approach is the effective cutoff, $\Lambda$, (introduced earlier) above which the modes of the theory under examination are regularized. The physics at the effective scale is encapsulated by the Wilsonian effective action, $S_\Lambda$, whose evolution with $\Lambda$ is given by the ERG equation. It is curious that, despite the success of the ERG in addressing
nonperturbative problems in Quantum Field Theory (QFT) (see~\cite{B+B,Wetterich-Rev,JMP-Review,Gies-Rev,Fisher-Rev,Aoki-Rev,Polonyi-Rev} for reviews) and despite the fact
that some of the most penetrating insights into supersymmetric theories utilize the Wilsonian effective
action (including the nonrenormalization theorems~\cite{nrts} and the Seiberg-Witten solution~\cite{S+W1,S+W2}) applications of the ERG to supersymmetric theories are rather limited, both in number and in scope~\cite{Bonini+Vian,Falkenberg,Bilal,Yoshida-SUSY04-1,Yoshida-SUSY04-2,Yoshida-SUSY07,Sonoda_WZ,Higashijima:2002mh,Higashijima:2003ki,Higashijima:2003rp,Higashi:2007tn,Higashi:2007dm} (see also the note added at the end of the paper). It
is hoped, then, that the concrete results that this paper provides will lead to a development
of this---surely fruitful---area.

The rest of this paper is arranged as follows. 
In \sec{Flow} we will discuss generalized ERGs and adapt the formalism to theories of a
scalar chiral superfield. Our subsequent analysis is facilitated by the introduction, in \sec{WEA}, 
of a  form for the Wilsonian effective action
in  which all the superspace coordinates are Fourier transformed. This allows us to directly
develop a simple diagrammatic representation for the flow equation, which is done
in \sec{diags}, and to prove the nonrenormalization theorem, which is the subject of \sec{NRT}.
In \sec{dual}, a construction is introduced (the `dual action' of~\cite{Trivial}) which is necessary for the analysis of
the existence of critical fixed points (\sec{CFPs}) and aids the discussion on the $\beta$-function
of the Wess-Zumino model (\sec{beta}). We conclude in \sec{conc}.

\begin{acknowledgments}
 I would like to thank IRCSET for financial support.
\end{acknowledgments}

\section{The Flow Equation}
\label{sec:Flow}

\subsection{Generalized ERG Equations}

Throughout this paper, we will work in $d=4$ Euclidean space. We will generally use the same
symbol for four-vectors and their moduli, the meaning hopefully being clear from the context.
In \app{Conventions} we review the approach of~\cite{Lukierski} to the problem of implementing Euclidean $\mathcal{N}=1$ superfields, and set our conventions. These conventions are such that one will get the correct signs when doing spinor algebra by using the appropriate formulae of Wess and Bagger~\cite{W+B}, but replacing the Minkowski metric by $\delta_{\mu\nu}$. Digging inside, however, there are some differences---notably in the definition of $\sigma_\mu$---but these can largely be forgotten about. Note that Hermitian conjugation is replaced by `Osterwalder and Schrader'
conjugation, which we will denote by OSC (schematically, for what we will do, this makes no difference).

Working, for the moment, in some generic QFT with fields $\varphi$, a generalized ERG follows from
the fundamental requirement that the partition 
function is invariant under the flow~\cite{WegnerInv,TRM+JL}:
\be
\label{eq:blocked}
-\flow e^{-S_\Lambda[\varphi]} =  \int_x \fder{}{\varphi(x)} \left(\Psi_x[\varphi] e^{-S_\Lambda[\varphi]}\right),
\ee
this property being ensured by 
the total derivative on
the right-hand side. 
The $\Lambda$ derivative
is performed at constant $\varphi$.
The functional, $\Psi$, parametrizes the continuum version of
a general Kadanoff blocking~\cite{Kadanoff}. 
To generate the family of flow equations to which 
Polchinski's formulation~\cite{pol} of the ERG belongs,
we take:
\be
	\label{eq:Psi}
	\Psi_x = \hf \dd^{\varphi \varphi}(x,y) \fder{\Sigma_\Lambda}{\varphi(y)},
\ee
where it is understood that we sum over all the elements
of the set of fields $\varphi$. The $\dd$s are the ERG kernels, 
which are generally
different for each of the elements of $\varphi$. 
In momentum space, each kernel
incorporates a cutoff function, $c(p^2/\Lambda^2)$, which 
dies off sufficiently fast as $p^2/\Lambda^2 \rightarrow \infty$ to
implement ultraviolet
regularization. The dot on top of the $\Delta$ is defined according to
\[
	\dot{X} \equiv -\flow X.
\]

Returning to~\eq{Psi}, and henceforth dropping the various subscripted $\Lambda$s,
we  take
\be
	\Sigma \equiv S - 2\hS,
\label{eq:Seed}
\ee
where $\hS$ is the `seed action'~\cite{aprop,scalar2,mgierg1,mgierg2}, a nonuniversal input
which controls the flow but of which all physical quantities should be independent. Given the
choice~\eq{Psi}, and a choice of cutoff function, the seed action encodes the residual blocking freedom. The only restrictions on the seed action are that it is infinitely differentiable and leads to convergent loop integrals~\cite{aprop,scalar2}. The first requirement is that of `quasi-locality' (mentioned in the introduction), which must apply to all
ingredients of the flow equation. Quasi-locality ensures that each ERG step is free
of IR divergences or, equivalently, that blocking is performed only over a local patch.
The seed action has the same structure and symmetries as the Wilsonian effective action; however, we choose the former, whereas we solve for the latter. Our flow equation reads:
\be
	-\flow  S 
	= \hf \fder{S}{\varphi} \cdot \dd \cdot \fder{\Sigma}{\varphi} 
	- \hf \fder{}{\varphi} \cdot \dd \cdot \fder{\Sigma}{\varphi}
\label{eq:ProtoFlow}
\ee
where, as ususal, we employ the shorthand
$A \cdot B \equiv \Int{x} A(x) B(x)$. Similarly, $A \cdot \dd \cdot B \equiv \int_{x,y} A_x \dd(x,y) B_y = \Int{p}/(2\pi)^{D} A(p) \dd (p) B(-p)$.
The two terms on the \rhs\ of~\eq{ProtoFlow} are often referred to as the classical and quantum terms, respectively, for reasons that will become apparent when we discuss the diagrammatics.

At this point, an example is useful. Suppose that we take $\varphi$ to be a single scalar field and
make the choice
\be
	\Delta(p) = \frac{c(p^2/\Lambda^2)}{p^2}.
\label{eq:Delta}
\ee
We interpret $\Delta(p)$ as a UV regularized or `effective' propagator.
Using this definition, we split the actions according to
\be
	S[\varphi] =  \hf \varphi \cdot \Delta^{-1} \cdot \varphi + \SR_\Lambda[\varphi],
	\qquad
	\hS[\varphi] = \hf \varphi \cdot \Delta^{-1} \cdot \varphi + \hSR_\Lambda[\varphi].
\label{eq:split}
\ee
These latter two expressions serve as a definition for what we mean by $\SR[\varphi]$
and $\hSR[\varphi]$; clearly, they can be interpreted as the interaction parts
of the Wilsonian effective action and seed action, respectively. Note that, just because we
have not included a mass term in the effective propagator, \eq{Delta}, does not necessarily mean that
the theory is massless: a mass term could be included in, or generated by, $\SR[\varphi]$. Thus~\eq{split} should be viewed simply as a convenient way of splitting the actions.

If we now substitute~\eqs{Delta}{split}
into~\eq{ProtoFlow} we get, up to a discarded vacuum energy term coming from the quantum term:
\be
	-\flow \SR = \hf \fder{\SR}{\varphi} \cdot \dd \cdot \fder{\SigmaR}{\varphi} 
	- \hf \fder{}{\varphi} \cdot \dd \cdot \fder{\SigmaR}{\varphi}
	- \varphi \cdot \Delta^{-1} \cdot \dd \cdot \fder{\hSR}{\varphi}.
\label{eq:Pol}
\ee
Note that all (non-vacuum) terms involving explicit $\Delta^{-1}$s, besides the final term which depends
on the interaction part of the seed action,
have cancelled amongst themselves; this observation
will be important when we come to construct an ERG for theories of a scalar chiral superfield.
If we were to set the interaction part of the seed action to zero---as we are quite at liberty to do---then the resulting equation is none other than Polchinski's form of the ERG equation. 

Ideally, since universal results must be independent of the choice of seed action,
we would like to retain a general seed action for all calculations. Unfortunately,
the methodology for the work pertaining to the (non) existence of fixed points has only been figured
out for the simplest seed action ($\hSR[\varphi]=0$). For other calculations
in this paper, however, we are able to keep a general seed action and will do so.

\subsection{An ERG for Theories of Scalar Chiral Superfields}

In this section, we construct an ERG for theories of a scalar chiral superfield. For most of this paper,
we will not consider any particular theory (\ie\ bare action) but rather will take the space of all
possible (quasi-local) theories as our arena: in other words, we consider all (quasi-local) theories of a scalar chiral superfield, $\suf$, and its conjugate, $\asuf$.  This is the correct setting for asking the question as to whether or
not there are any non-trivial fixed point theories. Only in \sec{beta} will we look at a specific theory---the Wess-Zumino model. 

\subsubsection{General Formulation}
\label{sec:general}

In the case of theories of a scalar chiral superfield, we find it convenient to automatically satisfy the chirality
constraint by taking the set of fields represented by $\varphi$ to be `potential superfields' (see \eg~\cite{WeinbergIII}), $\ptlsf$ and $\ptlasf$, which are related to the scalar chiral superfield, $\suf$, and its conjugate, $\asuf$, as follows:
\be
	\suf = \Dbar^2 \ptlsf , \qquad  \asuf = D^2 \ptlasf.
\label{eq:Ptl-SF}
\ee
In condensed notation, our flow equation reads:
\begin{multline}
		-\flow S =  \hf 
			\left(
			\fder{S}{\ptlasf} \cdot \dd^{\ptlasf \ptlsf} \cdot \fder{\Sigma}{\ptlsf}
			+ \fder{S}{\ptlasf} \cdot \dd^{\ptlasf \ptlasf} \cdot \Dbar^2 \cdot \fder{\Sigma}{\ptlasf}
			-\fder{}{\ptlasf} \cdot \dd^{\ptlasf \ptlsf} \cdot \fder{\Sigma}{\ptlsf}
			- \fder{}{\ptlasf} \cdot \dd^{\ptlasf \ptlasf} \cdot \Dbar^2 \cdot \fder{\Sigma}{\ptlasf}
			\right) 
	\\	
		+ \mathrm{OSC},
\label{eq:FlowEqn}
\end{multline}
where we have anticipated that it is convenient to extract a $\Dbar^2$ from the $\ptlasf\ptlasf$
kernel.
To be more explicit about what the dots mean in~\eq{FlowEqn}, we expand \eg
\be
	\fder{}{\ptlasf} \cdot \dd^{\ptlasf \ptlsf} \cdot \fder{}{\ptlsf} = 
	\FourInt{x} \FourVol{x'} \FourVol{\theta} \FourVol{\theta'}
	\fder{}{\ptlasf(x,\theta,\theta')} \dd^{\ptlasf \ptlsf}(x,\theta,\thetabar;x',\theta',\thetabar') \fder{}{\ptlsf(x,\theta,\theta')}.
\label{eq:dotNotation}
\ee
Given the superspace operators, $Q$ and $\anti{Q}$ [see~\eqs{Q}{Qbar}], supersymmetry of the flow equations follows straightforwardly, by considering the transformation
$\delta_\zeta \ptlsf = (\zeta Q + \overline{\zeta} \overline{Q}) \phi$, so long as we recognize that
\[
	 \dd^{XY}(x,\theta,\thetabar;x',\theta',\thetabar') =  \dd^{XY}(x-x',\theta-\theta',\thetabar-\thetabar'),
\]
where $X$ and $Y$ can each be either the potential superfield or its conjugate.

For what follows, including the development of a diagrammatic representation of the flow
equation, it is useful to work in completely Fourier transformed superspace; \ie\ we transform the fermionic coordinates as well as the spatial ones. 
Focussing first on the spatial coordinates,  we have the usual definitions:
\begin{align}
 	\ptlsf(x,\theta,\thetabar) = \MomInt{4}{p} \ptlsf(p,\theta,\thetabar) e^{-ip \cdot x},
	\qquad
	\ptlasf(x,\theta,\thetabar) = \MomInt{4}{p} \ptlasf(p,\theta,\thetabar) e^{ip \cdot x},
\\
	\dd^{XY} (x,\theta,\thetabar;x',\theta',\thetabar')
	=
	\MomInt{4}{p} 
	\dd^{XY} (p;\theta,\thetabar,\theta',\thetabar') e^{ip \cdot (x-x')}.
\label{eq:kernel}
\end{align}
The fermionic Fourier transforms are defined as follows:
\be
\label{eq:FFT}
	\ptlsf(p,\theta,\thetabar) =  4 \FermInt{4}{\rho} e^{-i\rho \cdot \theta }
						\ptlsf(p,\rho,\rhobar),
\qquad
	\ptlsf(p,\rho,\rhobar) = 4 \FermInt{4}{\theta} e^{i\rho \cdot \theta}
						\ptlsf(p,\theta,\thetabar),
\ee
where $\rho \cdot \theta \equiv \rho \theta+ \rhobar \thetabar$.
That we choose a factor of four to accompany both the Fourier transform and its inverse is a matter of convention. Indeed, any choice of prefactors whose product is sixteen would be consistent, as is apparent from~\eq{FermDeltaFn}.

When we completely Fourier transform the flow equation, \eqn{dotNotation} becomes:
\be
	\MomInt{4}{p} \FermInt{4}{\rho}
	\fder{}{\ptlasf(p,\rho,\rhobar)} \dd^{\ptlasf \ptlsf}(p) \fder{}{\ptlsf(p,\rho,\rhobar)},
\label{eq:FE-pasf-psf}
\ee
where we write
\be
	\dd^{\ptlasf \ptlsf}(p,\theta,\thetabar,\theta',\thetabar')
	=\dd^{\ptlasf \ptlsf}(p) \delta^{(4)}(\theta-\theta').
\ee
For the terms in the flow equation involving explicit $\Dbar^2$s or $D^2$s
we define
\begin{subequations}
\begin{align}
	\Dbar^2(p,\rho,\rhobar,\kappa,\anti{\kappa})  &  \equiv
	16 \FermInt{4}{\theta}
	e^{-i \rho \cdot \theta}
	\Dbar^2(p,\theta,\thetabar)
	e^{-i \kappa \cdot \theta}
\label{eq:D^2}
\\
\nonumber
	& =
		4p^2\fermprod{(\rhobar + \kappabar)}{(\rhobar + \kappabar)}
		-4\fermprod{\rhobar}{\rhobar} ((\rho+\kappa)p\kappabar)
		+4 \fermprod{\kappabar}{\kappabar} ((\rho+\kappa)p\rhobar)
\\
	&\qquad
		-\fermprod{\kappabar}{\kappabar} \fermprod{\rhobar}{\rhobar}
		\fermprod{(\rho + \kappa)}{(\rho + \kappa)},
\label{eq:FT_D}
\end{align}
\end{subequations}
and so arrive at the following building block of the flow equation:
\be
		\MomInt{4}{p} \FermInt{4}{\rho} \FermInt{4}{\kappa}
	\fder{}{\ptlsf(-p,\rho,\rhobar)} \dd^{\ptlsf \ptlsf}(p) 
	D^2(p,\rho,\rhobar,\kappa,\anti{\kappa}) 
	\fder{}{\ptlsf(p,\kappa,\anti{\kappa})}.
\label{eq:FE-psf-psf}
\ee

Our aim now is to mimic the decomposition~\eq{split}. To this end, we write
\be
	S[\ptlasf,\ptlsf] = -\ptlasf \cdot D^2 \cdot c^{-1} \cdot \Dbar^2 \cdot \ptlsf
		-2m_0 \ptlsf \cdot c^{-1} \cdot \Dbar^2 \cdot \ptlsf
		-2m_0 \ptlasf \cdot c^{-1} \cdot D^2 \cdot \ptlasf
		+\SR[\ptlasf,\ptlsf],
\label{eq:SUSY-split}
\ee
where $m_0$ is the bare mass. Actually, as a consequence of the nonrenormalization
theorem, the mass is the same at all scales and so there is no need to call it the bare mass.
However, we will shortly perform some rescalings, after which the superpotential will
renormalize, via the scaling dimension of the field. In this case, it will be useful to distinguish
the bare mass from the running mass.

It is worth pointing out that, in contrast to the case of plain scalar field theory, we find
it convenient to pull out the mass terms from $\SR$. As we will see below, the
reason for this is because, unlike $\Delta^{\ptlasf \ptlsf}$ (or the effective propagator in
scalar field theory), $\Delta^{\ptlsf \ptlsf}$ vanishes for $m_0 = 0$.

Note that, since we include a momentum dependent cutoff function in the two-point $\ptlsf\ptlsf$
vertex, this term contributes to both the superpotential and the K\"{a}hler potential, as can be
seen by expanding $c(p^2/\Lambda^2) = 1 + \order{p^2/\Lambda^2}$. If we now make the following (very natural~\cite{1001}) choices
for the momentum space integrated ERG kernels
\begin{subequations}
\begin{align}
	\Delta^{\ptlasf \ptlsf}(p) & = \frac{1}{16} \frac{c(p^2)}{p^2+m_0^2},
\label{eq:EP-asf-sf}
\\
	\Delta^{\ptlsf \ptlsf}(p)  & =  \Delta^{\ptlasf \ptlasf}(p)  = \frac{1}{64} \frac{m_0 c(p^2)}{p^2(p^2+m_0^2)},
\label{eq:EP-sf-sf}
\end{align}
\end{subequations}
then we once again find that the only place where the explicitly written two-point
terms in~\eq{SUSY-split} appear is in a term containing the seed
action [\cf~\eq{Pol}]:
\begin{multline}
	-\flow \SR = 
		\left(
			\ptlsf \cdot \Dbar^2 + 4m_0 \ptlasf
		\right)  \cdot c^{-1} \cdot D^2 \cdot
		\left(
			\dd^{\ptlasf\ptlsf} \cdot \fder{\hSR}{\ptlsf}
			+\dd^{\ptlasf\ptlasf} \cdot \Dbar^2 \cdot \fder{\hSR}{\ptlasf}
		\right)
	\\
		+\hf 
		\left(
			\fder{\SR}{\ptlasf} \cdot \dd^{\ptlasf \ptlsf} \cdot \fder{\SigmaR}{\ptlsf}
			+ \fder{\SR}{\ptlasf} \cdot \dd^{\ptlasf \ptlasf} \cdot \Dbar^2 \cdot \fder{\SigmaR}{\ptlasf}
			-\fder{}{\ptlasf} \cdot \dd^{\ptlasf \ptlsf} \cdot \fder{\SigmaR}{\ptlsf}
			- \fder{}{\ptlasf} \cdot \dd^{\ptlasf \ptlasf} \cdot \Dbar^2 \cdot \fder{\SigmaR}{\ptlasf}
		\right)
	\\
		+
		\mathrm{OSC}.
\label{eq:explicit}
\end{multline}

Setting $\hSR=0$ yields the supersymmetric version of Polchinski's equation.
Deriving~\eq{explicit} is, however, somewhat more involved than in the case of scalar field theory, due
to the fact that the two-point K\"{a}hler vertex is not invertible. 
Nevertheless, we do have at our
disposal the relationship
\be
D^2 \Dbar^2 D^2 = -16p^2 D^2,
\label{eq:DDD}
\ee
and it is this is which ensures that everything goes through. 

It is tempting to identify the integrated kernels as regularized propagators, but we must be careful doing so. In scalar field theory, it is both natural and
convenient to make this identification. However, in the current case we cannot invert the kinetic term, and so it is not immediately obvious that we can define a propagator.

This situation is somewhat similar to what occurs in the manifestly gauge invariant
ERGs for QCD~\cite{qcd} and QED~\cite{qed} where, again, the two-point vertex cannot be inverted.
The central point is that  ERG kernels
exist, first and foremost, as ingredients of a perfectly well defined ERG equation, and there
is nothing to stop us from integrating them. If it so happens that one can \emph{additionally} identify
the integrated kernels with regularized propagators then all the better, but this occurs only in special cases and not for general field content. Nevertheless, even when this identification cannot be made, the integrated kernels have a structural similarity to regularized propagators 
and play an analogous role in ERG diagrams to the
role played by normal propagators in Feynman diagrams. With this in mind, the phrase `effective propagator' was coined~\cite{aprop}.

In the current scenario, things are somewhere between the case of scalar theory and
manifestly gauge invariant formulations. As emphasised by Weinberg~\cite{WeinbergIII} (chapter 30),
the theory is invariant under the `gauge' transformations
\[
	\ptlsf \rightarrow \ptlsf +  \Dbar_{\alphadot} \omegabar^{\alphadot}, 
	\qquad \ptlasf +  \ptlasf \rightarrow D^\alpha \omega_\alpha,
\]
where $\omega$ and its conjugate are unconstrained superfields. This invariance comes about
because the theory is built out of gauge invariant objects, $\suf$ and $\asuf$, in contrast to
gauge theories where the theory is built using the gauge \emph{variant} connection.
Now, in the context of theories of a scalar chiral superfield, 
so long as one is only interested in correlation functions of gauge invariant objects,
then one can proceed without fixing the gauge by introducing new variables of
integration in the path integral. This involves separating out the zero mode
of the two-point operator~\cite{WeinbergIII}. The resulting propagators are
(modulo the UV regularization) precisely what we obtain for the integrated ERG
kernels. Thus, with this understanding, we can interpret the integrated ERG kernels
as regularized propagators.

Returning to~\eq{explicit}, it is
worth adding that, reassuringly
in this supersymmetric scenario, the vacuum terms vanish.

\subsubsection{Rescalings}
\label{sec:rescale}

One of the applications of our flow equation will be to analyse the existence of fixed points.
Fixed point behaviour is most easily seen by rescaling to dimensionless variables, by
dividing by $\Lambda$ to the appropriate scaling dimension (by this it is meant, of course, the full scaling dimension, and not the canonical dimension). As it turns out, there is a subtlety related to
scaling
out the anomalous dimension from $\ptlsf$ (and $\ptlasf$), so we will consider this rescaling first, in isolation.
Thus, we make the following transformation:
\be
	\ptlsf \rightarrow \ptlsf  \sqrt{Z}, \qquad \ptlasf \rightarrow \ptlasf  \sqrt{Z}
\label{eq:rescale}
\ee
where $Z$ is the field strength renormalization, from which we define the anomalous dimension:
\be
	\gamma \equiv \Lambda \der{\ln Z}{\Lambda}.
\label{eq:eta}
\ee
The problem with this transformation is that it produces an annoying factor of $1/Z$ on the
\rhs\ of the flow equation. However, we can remove this factor by utilizing the immense freedom
inherent in the ERG, encapsulated by~\eq{blocked}, to shift the kernels 
$\dd^{XY} \rightarrow Z \dd^{XY}$. For orientation, the resulting flow equation is therefore not obtainable from the Polchinski equation by a simple rescaling of the fields: it is a cousin, rather than a descendent. 
In the case of scalar field theory, such a flow equation (with $\hSR=0$) was first considered in~\cite{Ball}; the version with more general seed action has been considered in~\cite{scalar1,scalar2}.

With this change to the flow equation, \eq{FlowEqn} becomes:
\begin{multline}
	-\flow S + \frac{\gamma}{2} 
		\left(\ptlsf \cdot \fder{S}{\ptlsf} + \ptlasf \cdot \fder{S}{\ptlasf} \right)
	\\
			 =  \hf 
			\left(
			\fder{S}{\ptlasf} \cdot \dd^{\ptlasf \ptlsf} \cdot \fder{\Sigma}{\ptlsf}
			+ \fder{S}{\ptlasf} \cdot \dd^{\ptlasf \ptlasf} \cdot \Dbar^2 \cdot \fder{\Sigma}{\ptlasf}
			-
			\fder{}{\ptlasf} \cdot \dd^{\ptlasf \ptlsf} \cdot \fder{\Sigma}{\ptlsf}
			- \fder{}{\ptlasf} \cdot \dd^{\ptlasf \ptlasf} \cdot \Dbar^2 \cdot  \fder{\Sigma}{\ptlasf}
			\right) + \mathrm{OSC}.
\label{eq:Flow}
\end{multline}
Note that, as a consequence of our rescalings, the superpotential does now renormalize, but only through the field strength renormalization.

We now complete the rescalings
started with~\eq{rescale}. To this end, we define the `RG-time'
\be
	t \equiv \ln \mu/\Lambda,
\ee
where $\mu$ is an arbitrary mass scale, and  also scale out
the various canonical dimensions:
\be
	\qquad p_i \rightarrow p_i \Lambda,
	\qquad \rho_i \rightarrow \rho_i \sqrt{\Lambda}.
\label{eq:Canonical}
\ee
In these units, fixed point solutions satisfy the condition
\be
	\partial_t S_\star[\ptlasf,\ptlsf] = 0.
\label{eq:FP}
\ee
This follows because, if all variable are measured in terms of $\Lambda$, independence of $\Lambda$ implies
scale independence. (Subscript $\star$s will be used to denote fixed-point quantities.) 

With these rescalings,  the flow equation in the massless case reads
\be
	\left[
		\partial_t 
		+ \frac{\gamma}{2} 
		\left(
			\ptlsf \cdot \fder{}{\ptlsf} + \ptlasf \cdot \fder{}{\ptlasf} 
		\right)
		+ \frac{\Delta_D}{2} -2
	\right] S
	=
	\frac{1}{16}
	\left(
		\fder{S}{\ptlasf} \cdot c' \cdot \fder{\Sigma}{\ptlsf}
		- \fder{}{\ptlasf} \cdot c' \cdot \fder{\Sigma}{\ptlsf}
	\right) + \mathrm{OSC},
\label{eq:Rescaled}
\ee
where, with $p$ now being dimensionless,
\[
	c'(p^2) \equiv \pder{}{p^2} c(p^2),
\]
and the `superderivative counting operator', $\Delta_D$, (utterly unrelated to the effective propagator, $\Delta$) is given by
\bea
\nonumber
	\Delta_D & \equiv & 
	2
	\left[
		-2 + \MomInt{4}{p} \FourVol{\rho} \ptlsf(p,\rho,\rhobar)
		\left(
			p^\mu \pder{'}{p^\mu} + \hf \rho^\alpha \pder{}{\rho^\alpha}
			+ \hf \rhobar_{\dot{\alpha}} \pder{}{\rhobar_{\dot{\alpha}}}
	\right)
	\fder{}{\ptlsf(p,\rho,\rhobar)}
	\right.
\\&&
	\qquad
	\left.
		+ \MomInt{4}{p} \FourVol{\rho} \ptlasf(p,\rho,\rhobar)
		\left(
			p^\mu \pder{'}{p^\mu} + \hf \rho^\alpha \pder{}{\rho^\alpha}
			+ \hf \rhobar_{\dot{\alpha}} \pder{}{\rhobar_{\dot{\alpha}}}
		\right)
		\fder{}{\ptlasf(p,\rho,\rhobar)}
	\right].
\label{eq:counting}
\eea
The prime on the momentum derivative, \viz\ $\partial' /\partial p_\mu$, means that the derivative is not allowed to
strike the momentum conserving $\delta$-function which belongs to each vertex.
The flow equation~\eq{Rescaled} generalizes the dimensionless flow equation of scalar field theory~\cite{Ball,B+B,TRM-Deriv}, in an obvious way.

Finally, in anticipation of our study of the $\beta$-function of the Wess-Zumino model (\sec{beta}), it is convenient to return to the flow equation~\eq{Flow}. Studying the $\beta$-function is a different
problem from looking for the complete spectrum of fixed point theories and  
there is nothing to be gained by scaling the out the various canonical dimensions.
However, in this context, it is worth rescaling the fields by the three-point, superpotential coupling, $\lambda$: $\ptlsf \rightarrow \ptlsf/\lambda$ (and similarly for $\ptlasf$). By doing so, the perturbative
expansion in $\lambda^2$ coincides with the one in $\hbar$, and this is the natural way to
do perturbation theory (of course, there is no absolute need to perform this rescaling, but it does
make life somewhat easier if we do so).

We absorb the change on the \lhs\ of the flow equation resulting from this rescaling into the term involving the anomalous dimension, $\gamma$. With this latter rescaling, the perturbative expansion of the action, should we choose to perform one, reads:
\be
	S \sim \sum_{i=0}^\infty \lambda^{2(i-1)} S_i,
\label{eq:Pert-action}
\ee
where $S_0$ is the classical action, and the $S_{\geq1}$ are the quantum corrections. 

The flow equation in the current scenario reads:
\begin{multline}
		-\flow S + \frac{\tilde{\gamma}}{2} 
		\left(\ptlsf \cdot \fder{S}{\ptlsf} + \ptlasf \cdot \fder{S}{\ptlasf} \right)
	\\
		 =  \hf 
		\left(
		\fder{S}{\ptlasf} \cdot \dd^{\ptlasf \ptlsf} \cdot \fder{\Sigma_\lambda}{\ptlsf}
		+ \fder{S}{\ptlasf} \cdot \dd^{\ptlasf \ptlasf} \cdot \Dbar^2 \cdot \fder{\Sigma_\lambda}{\ptlasf}
		- \fder{}{\ptlasf} \cdot \dd^{\ptlasf \ptlsf} \cdot \fder{\Sigma_\lambda}{\ptlsf}
		- \fder{}{\ptlasf} \cdot \dd^{\ptlasf \ptlasf} \cdot \Dbar^2 \cdot \fder{\Sigma_\lambda}{\ptlasf}
		\right) 
	\\	
		+ \mathrm{OSC}
\label{eq:Flow_2}
\end{multline}
where\footnote{Note that, in contrast to some other works~\cite{scalar2,aprop}, we have pulled a $\lambda^2$ out of the seed action, as well as the Wilsonian effective action.}
\be
	\Sigma_\lambda = \lambda^2 (S - 2\hS).
\label{eq:Sigma_l}
\ee

\section{The Wilsonian Effective Action}
\label{sec:WEA}

As emphasised throughout this paper, most of the time we will consider a general (quasi-local) theory of scalar chiral superfields. This means that, apriori, the superpotential and the \Kahler\ potential possess all possible interactions. Expanding the action in powers of the fields, the superpotential possesses a two-point vertex with coupling $f^{(2)}$, a three-point vertex with coupling $f^{(3)}$, and so forth. (Note that
we can choose to exclude one-point vertices in the superpotential through a classical renormalization condition: there are no quantum corrections as a consequence of the nonrenormalization theorem.)
With this in mind, we write the superpotential as
\begin{align}
	\nonumber
	\SP[\ptlsf] & =\sum_{n=2}^\infty \frac{\SP^{(n)}}{n!}
		\FourInt{x} 
		\FourVol{\theta}
		\delta^{(2)}(\thetabar)
		\left[
			\prod_{j=1}^n \suf(x,\theta,\thetabar)
		\right]
	\\
		& =
		-4 \sum_{n=2}^\infty \frac{\SP^{(n)}}{n!}
		\FourInt{x} \FourVol{\theta} \ptlsf(x,\theta,\thetabar)
		\left[
			\prod_{j=2}^n
			\Dbar^2(x,\theta,\thetabar) \ptlsf(x,\theta,\thetabar)
		\right].
\label{eq:SPtl}
\end{align}

We will take Weinberg's definition~\cite{WeinbergIII}
of the \Kahler\ potential: it is a real scalar function of $\suf$ and $\asuf$,
where we allow terms in which  superderivatives act on these fields.
(Sometimes the \Kahler\ potential is defined to be just the piece without
any additional superderivatives.) The \Kahler\ potential is a sum over all
terms with  $n$ $\asuf$s and $m$ $\suf$s. Actually, we choose to write the \Kahler\
potential in terms of the potential superfield, $\ptlsf$, and its conjugate.
Suppressing superspace coordinates,
the vertex for the $n$, $m$ contribution is $K^{(n,m)}$, where this object is understood to be a differential operator,
containing at least $n$ $D^2$s and $m$ $\Dbar^2$s:
\begin{align}
	\Kah[\ptlasf,\ptlsf] & =
		- \sum_{n+m\geq2}^\infty \frac{4^{2-n-m}}{n! m!}
		\left[
			\prod_{j=0}^n \FourInt{x_j} \FourVol{\theta'_j} \ptlasf(x'_j,\theta'_j,\thetabar'_j)
		\right]
		\left[
			\prod_{k=0}^m \FourInt{x_k} \FourVol{\theta_k} \ptlsf(x_k,\theta_k,\thetabar_k)
		\right]
	\nonumber
	\\
		& 
		\qquad
		{\Kah}^{(n,m)}(x'_1,\ldots,x'_n,\theta'_1,\ldots,\theta'_n, \thetabar'_1,\ldots,\thetabar'_n;
		x_1,\ldots,x_m,\theta_1,\ldots,\theta_m, \thetabar_1,\ldots,\thetabar_m).
\label{eq:Kahler}
\end{align}
In this expression, the operator
$K^{(n,m)}$ acts to its left (to save space!). 
Notice that we use primed coordinates for $\ptlasf$s and unprimed coordinates for $\ptlsf$s. The factor of $4^{2-n-m}$ is inserted for later convenience.
Although not manifest in the way we have written things,
every vertex implements locality in the fermionic coordinates.
For small numbers of fields, we will often use a notation where the fields are indicated, explicitly \eg\ $K^{\ptlasf \ptlsf} \equiv K^{(1,1)}$.

The fermionic Fourier transforms of the vertices follow from substituting~\eq{FFT} and its conjugate into~\eqs{Kahler}{SPtl}. 
Let us start by considering the completely Fourier transformed
superpotential. To cast this in a neat form, we use a trick.
In the second line of~\eq{SPtl}, we pretend that there are $n$
different $\theta$s. Each field (and each $\Dbar^2$) is taken to
depend on one of these $\theta_i$ and its conjugate. With this
in mind, we rewrite
\begin{align*}
	\FourInt{\theta} 
	&= \FourInt{\theta_1} \cdots \FourInt{\theta_n}
	\delta^{(4)}(\theta_1-\theta_2) \cdots \delta^{(4)}(\theta_{n-1}-\theta_n) 
\\
	&= 
	16^{n-1}
	\FourInt{\theta_1} \cdots \FourInt{\theta_n}
	 \FourInt{\omega_{12}} \cdots \FourInt{\omega_{n-1\, n}}
	e^{i\omega_{12} \cdot (\theta_1-\theta_2)}
	\cdots
	e^{i\omega_{n-1 \, n} \cdot (\theta_{n-1}-\theta_n)},
\end{align*}
where we have used the representation of the Fermionic $\delta$-function,
\eq{FermDeltaFn}. Using~\eq{D^2}, it is now a simple matter to check that
\begin{multline}
	f[\ptlsf] =  -4 \sum_{n=2}^{\infty} \frac{4^{n-2} f^{(n)}}{n!}
	\left[
		\prod_{j=1}^n \MomInt{4}{p_j} \FourVol{\rho_j} 
		\ptlsf(p_j,\rho_j,\rhobar_j)
	\right]
	\hat{\delta}(p_1+\cdots+p_n)	
\\
	\left[
	\prod_{i=2}^{n}
	\FourInt{\omega_{i-1\,i}}
	\Dbar^2(p_i,\omega_{i-1\, i} - \omega_{i\, i+1}, \anti{\omega}_{i-1\, i} - \anti{\omega}_{i\, i+1},
	\rho_i,\rhobar_i)
	\right]
	\delta^{(4)}(\omega_{12}-\rho_1),
\label{eq:FTSP}
\end{multline}
where $\omega_{n\,n+1} \equiv 0$ (and similarly for its conjugate)
and we have introduced the notation
\[
\hat{\delta}(p) \equiv (2\pi)^4 \delta^{(4)}(p).
\] 
Note that there is no conservation of the fermionic `momenta'. 
For the computation of $\beta$-function coefficients,
it will be useful to write this as
\be
	f[\ptlsf]  = - \sum_{n=2}^{\infty} \frac{1}{n!}
	\left[
		\prod_{j=1}^n \MomInt{4}{p_j} \FourVol{\rho_j} 
		\ptlsf(p_j,\rho_j,\rhobar_j)
	\right]
	\hat{\delta}(p_1+\cdots+p_n)	
	F^{(n)}(p_1,\rho_1,\rhobar_1;\ldots;p_n,\rho_n,\rhobar_n).
\label{eq:F^n}
\ee

We will find a similar structure to~\eq{FTSP} when we completely
Fourier transform the \Kahler\ potential.
To get a feeling for this, let us start by looking at the classical two-point vertices.
In position superspace, the two-point, classical contribution to the $\ptlasf \ptlsf$ vertex is given by
\be
	-\FourInt{x} \FourVol{x'} \FourVol{\theta} 
	c_\Lambda^{-1}(x-x') 
	D^2 \ptlasf(x,\theta,\thetabar)
	\Dbar^2 \ptlsf(x',\theta,\thetabar),
\label{eq:KCTP-pos}
\ee
where we recall that, in momentum space, $c(p^2/\Lambda^2)$ is a smooth ultraviolet cutoff function [see~\eq{kernel} for the definition of the Fourier transform], which regularizes the theory above the scale $\Lambda$. Since the only dependence of $D^2$ and $\Dbar^2$ on position coordinates occurs via spacetime derivatives, in Fourier transformed superspace we have:
\be
		{\Kah}_0^{\ptlasf \ptlsf}(-p,\rho,\rhobar;p,\kappa,\anti{\kappa}) = -16c^{-1}(p^2/\Lambda^2)
	\FermInt{4}{\theta}
	\left[
		D^2(-p,\theta,\thetabar) e^{i \rho\cdot \theta}
	\left]
	\left[
		\Dbar^2(p,\theta,\thetabar) e^{-i \kappa \cdot \theta}
	\right]
	\right.
	\right.
	\!\!\!,
\label{eq:KCTP-mom}
\ee
where the subscript `$0$' on the vertex indicates that we are considering only the classical contribution,
\eq{KCTP-pos}.
Notice that, if we were to integrate by parts in superspace, so as to transfer the $D^2$ from the $\ptlasf$ to the $\ptlsf$, then we should remember to change the argument $-p$ to $+p$. 
Applying~\eqs{D^2}{FermDeltaFn}, it is straightforward to show that~\eq{KCTP-mom} can be rewritten in the intuitive form:
\begin{subequations}
\begin{align}
	{\Kah}_0^{\ptlasf \ptlsf}(-p,\rho,\rhobar;p,\kappa,\anti{\kappa}) 
	& = 
	-c^{-1}(p^2/\Lambda^2)
	\FermInt{4}{\omega}
		D^2(-p,\omega,\anti{\omega},\rho,\rhobar)
		\Dbar^2(p,\omega,\anti{\omega},\kappa,\anti{\kappa}),
\label{eq:KCTP-mom2}
	\\
	& = 
	-c^{-1}(p^2/\Lambda^2)
	\FermInt{4}{\omega}
		D^2(p,\rho,\rhobar,\omega,\anti{\omega})
		\Dbar^2(-p,\kappa,\anti{\kappa},\omega,\anti{\omega}),
\label{eq:KCTP-mom2b}
\end{align}
\end{subequations}
where the last line, which will be useful later, follows from inspection of~\eq{FT_D}.
Contracting two such vertices into one another gives
\be
	\FermInt{4}{\omega} {\Kah}_0^{\ptlasf \ptlsf}(-p,\rho,\rhobar;p,\omega,\anti{\omega})
	{\Kah}_0^{\ptlasf \ptlsf}(-p,\omega,\anti{\omega};p,\kappa,\anti{\kappa}) 
	=+16p^2 c^{-1}(p^2/\Lambda^2) {\Kah}_0^{\ptlasf \ptlsf}(-p,\rho,\rhobar;p,\kappa,\anti{\kappa}),
\label{eq:KCTP-prod}
\ee
which is a manifestation of the superspace relationship~\eq{DDD}.

As mentioned earlier, since we include a cutoff function in the mass term, the classical, two-point mass vertices contribute to both the superpotential \emph{and} the \Kahler\ potential. In position space we have the contribution to the action
\be
	- \frac{1}{2!} 4m \FourInt{x} \FourVol{x'} \FourVol{\theta} c_\Lambda^{-1}(x-x')
		\left( \ptlsf \Dbar^2 \ptlsf + \ptlasf D^2 \ptlasf \right),
\label{eq:CTP-m}
\ee
where we have pulled out a factor of $1/2!$, in view of~\eqs{Kahler}{SPtl}. In completely Fourier transformed superspace, we have:
\begin{subequations}
\begin{align}
	{S}^{\ptlasf \ptlasf}_0(-p,\rho,\rhobar;p,\kappa,\anti{\kappa}) 
		& = -4m_0c^{-1}(p^2/\Lambda^2) D^2(p,\rho,\rhobar,\kappa,\anti{\kappa}),
\label{eq:CTP-m-sf-FT}
\\
		{S}^{\ptlsf \ptlsf}_0(p,\rho,\rhobar;-p,\kappa,\anti{\kappa})
	& = -4m_0c^{-1}(p^2/\Lambda^2) \Dbar^2(-p,\rho,\rhobar,\kappa,\anti{\kappa}).
\label{eq:CTP-m-asf-FT}
\end{align}
\end{subequations}
For completeness, we give the explicit expression for the completely Fourier transformed classical,
two-point vertices in \appx{CTP}.

Now we want to deal with general contributions to the \Kahler\ potential.
Noting that, if superfields carry positive momenta into the vertices, then anti-superfields carry positive momenta out of the vertices, we define the momentum space vertices (suppressing fermionic coordinates) via:
\begin{widetext}
\bea
\nonumber
	\lefteqn{
		\Kah^{(n,m)}(-p'_1,\ldots,-p'_n, \ldots;p_1,\ldots p_m,\ldots)\, \hat{\delta} \!
		\left(
			-\sum_{j=1}^n p'_j + \sum_{k=1}^m p_k
		\right)
	} \\
	& &
	\nonumber
	=
	\left(
		\prod_{i=1}^{n} \FourInt{x'_i}
	\right)
	\left(
		\prod_{j=1}^{m} \FourInt{x_{j}}
	\right)
	\Kah^{(n,m)}(x'_1,\ldots,x'_n,\ldots;x_1,\ldots, x_m,\ldots)
	\\
	&&
	\qquad
	\exp \!\!
	\left(
		i \sum_{k=1}^n p'_k \cdot x'_k - i \sum_{l=1}^m p_l \cdot x_l
	\right)
\eea
\end{widetext}
so that all momenta flow into the vertex coefficient functions. 

The idea now is to break up
$\Kah^{(n,m)}$ into $n$ pieces, denoted by $\Kah'^{(n,m)}_i(-p'_i,\theta'_i,\thetabar'_i)$, associated with the $\ptlasf(p'_i,\theta'_i,\thetabar'_i)$
and $m$ pieces, denoted by $\Kah^{(n,m)}_j(p_j,\theta_j,\thetabar_j)$, associated with the 
$\ptlsf(p_j,\theta_j,\thetabar_j)$.
These objects might, individually, possess loose spinor indices which are contracted
together in some way but, for brevity, we will not explicitly indicate this:
\begin{multline}
	K[\ptlasf,\ptlsf]
	=
	-  \sum_{n+m\geq2}^\infty \frac{4^{2-n-m}}{n! m!}
	\left[
		\prod_{j=0}^n \MomInt{4}{p'_j} \FourVol{\theta'_j} 
		K'^{(n,m)}_j(p'_j,\theta'_j,\thetabar'_j) \ptlasf(p'_j,\theta'_j,\thetabar'_j)
	\right]
\\
	\left[
		\prod_{k=0}^m \MomInt{4}{p_k} \FourVol{\theta_k} 
		K^{(n,m)}_k (p_k,\theta_k,\thetabar_k) \ptlsf(p_k,\theta_k,\thetabar_k)
	\right]
\\
	\delta^{(4)}(\theta'_1-\theta'_2) \cdots \delta^{(4)}(\theta_{m-1} - \theta_m)
	\hat{\delta} \!
		\left(
			-\sum_{j=1}^n p'_j + \sum_{k=1}^m p_k
		\right).
\label{eq:Kahler-FT}
\end{multline}
With this in mind,
we can obtain a particularly useful form for the completely Fourier transformed vertices by
generalizing~\eq{D^2}:
\be
	K^{(n,m)}_j (p,\rho,\rhobar,\kappa,\anti{\kappa})  \equiv
	16 \FermInt{4}{\theta_j}
	e^{-i \rho \cdot \theta_j}
	K^{(n,m)}_j (p,\theta_j,\thetabar_j)
	e^{-i \kappa \cdot \theta_j}.
\label{eq:K_j}
\ee

Using the same trick we used for completely Fourier transforming the
superpotential, we find:
\begin{multline}
	\Kah[\ptlasf,\ptlsf] = - \sum_{n+m\geq2}^\infty \frac{1}{n! m!}
\\
	\prod_{j=1}^n
	\MomInt{4}{p'_j} \FourVol{\rho'_j} 
	\FourVol{\omega'_{j\, j+1}}
	\;
		\ptlasf(p'_j,\rho'_j,\rhobar'_j)
	\;
	K'^{(n,m)}_j (-p'_j,\omega'_{j\, j+1} - \omega'_{j-1\, j},
		\anti{\omega}'_{j\, j+1} - \anti{\omega}'_{j-1\, j},\rho'_j,\rhobar'_j)
\\
	\prod_{i=1}^m
	\MomInt{4}{p_i} \FourVol{\rho_i} 
	\FourVol{\omega_{i-1\, i}}
	\;	
	\ptlsf(p_i,\rho_i,\rhobar_i)
	\;
	K^{(n,m)}_i (p_i,\omega_{i-1\, i} - \omega_{i\, i+1},
		\anti{\omega}_{i-1\, i} - \anti{\omega}_{i\, i+1},\rho_i,\rhobar_i)
\\
	\hat{\delta}(-p'_1-\cdots-p'_n+p_1+\cdots p_m)
	\delta^{(4)}(\omega_{01} - \omega'_{n\, n+1}),
\label{eq:FTK}
\end{multline}
where $\omega'_{01}, \ \omega_{m\,m+1} \equiv 0$ (and similarly for their conjugates).
Notice that the annoying factor of $4^{2-n-m}$ has disappeared.

We conclude this section with some remarks on the form of the $K^{\ptlasf\ptlsf}$
vertex, which we will require later. The vertex must possess at least one $D^2$ and at least one
$\Dbar^2$. The observation we will require is that general two-point vertices can be taken to
have only additional powers of momenta and no further superderivatives. To see this, we start
by noting that, as usual,
\begin{subequations}
\begin{align}
	\{D_\alpha, D_\beta\} = \{\Dbar_{\dot{\alpha}}, \Dbar_{\dot{\beta}} \} & = 0,
	\label{eq:D,D}
	\\
	\{D_\alpha, \Dbar_{\dot{\alpha}} \}  & =  -2 i \partial_{\alpha \dot{\alpha}}.
\label{eq:D,Dbar}
\end{align}
\end{subequations}
Since space-time derivatives can thus be written in terms of superderivatives, a general two-point
vertex goes like
\be
	\Dbar^2 \cdots D^2,
\label{eq:string}
\ee
where the ellipsis stands for an arbitrary string of superderivatives (with epsilon tensors included, 
as appropriate) and 
we have used integration by parts in superspace to arrange for all superderivatives to 
strike one of the fields. If the ellipsis represents unity, then our assertion is clearly satisfied.
Otherwise, we must have either
\[
	\Dbar^2 \cdots \Dbar_{\dot{\alpha}} \Dbar_{\dot{\beta}} D^2, \qquad \mathrm{or} \qquad
	\Dbar^2 \cdots D_\alpha \Dbar_{\dot{\alpha}} D^2.
\]
Dropping overall constants, we can use~\eq{D,D} to rewrite the first term and~\eq{D,Dbar} 
to rewrite the second, as follows:
\[
	\epsilon_{\dot{\alpha} \dot{\beta}} \Dbar^2 \cdots \Dbar^2 D^2,
	\qquad  \mathrm{or} \qquad
	p_{\alpha \dot{\alpha}} \Dbar^2 \cdots D^2.
\]
Iterating the procedure until the ellipses have been removed, we see
that a general two-point vertex can be written as a string of $D^2$s and
$\Dbar^2$s, up to powers of momentum. However, we can use the
relationship~\eq{DDD}
to reduce these strings to a single $D^2$ and a single $\Dbar^2$, up to
powers of momentum, thereby proving the original assertion.

\section{Diagrammatics}
\label{sec:diags}

One of the advantages of Fourier transforming all superspace coordinates is that the vertices
are converted from differential operators to functions. These vertices can thus be given a
straightforward diagrammatic interpretation. 
The diagrammatics for the action is most simply introduced by considering the two-point
vertex, ${S}^{\ptlasf \ptlsf}$:
\be
	{S}^{\ptlasf \ptlsf}(-p,\rho,\rhobar;p,\kappa,\anti{\kappa}) \equiv \ensuremath{\begin{array}{c}\input{pstex/CTP-KP.pstex_t} \end{array}}.
\label{eq:Diags-CTP}
\ee
The arrows on the lines emanating from the vertex indicate whether the corresponding fields
are potential superfields or potential anti-superfields.  We could instead have simply tagged each line with a $\ptlsf$ or $\ptlasf$, as appropriate. However, we have avoided doing this to emphasise that the diagrammatics involves only the vertex coefficient functions, the fields
and symmetry factors having been stripped off.  
To represent higher point vertices, we simply add more legs, as appropriate.
Usually, we will drop all coordinate labels, and arrows, for brevity.

The diagrammatic form of the various flow equations follows by direct substitution of the diagrammatic
form of the action  and identifying terms with the same field content.
Taking the flow equation~\eq{Flow_2}, for definiteness,
the result is shown in \fig{Flow}, where $\{f\}$ is a set of any $n_f$ $\ptlsf$s and/or $\ptlasf$s. Note that, since all fields have been stripped off, we can write the $\Lambda$-derivative as a total, rather than partial, derivative.
\bcf[h]
	\[
	\left(
		-\totalflow + \hf  \tilde{\gamma} n_f
	\right)
	\dec{
		\ensuremath{\begin{array}{c}\begin{picture}(0,0)%
\includegraphics{pstex/Vertex-S.pstex}%
\end{picture}%
\setlength{\unitlength}{3947sp}%
\begingroup\makeatletter\ifx\SetFigFont\undefined%
\gdef\SetFigFont#1#2#3#4#5{%
  \reset@font\fontsize{#1}{#2pt}%
  \fontfamily{#3}\fontseries{#4}\fontshape{#5}%
  \selectfont}%
\fi\endgroup%
\begin{picture}(341,318)(2180,-963)
\put(2291,-859){\makebox(0,0)[lb]{\smash{{\SetFigFont{11}{13.2}{\rmdefault}{\mddefault}{\updefault}{\color[rgb]{0,0,0}$S$}%
}}}}
\end{picture}%
 \end{array}}
	}{\{f\}}
	=
	\frac{1}{2}
	\dec{
		\ensuremath{\begin{array}{c}\input{pstex/Dumbbell-S-Sigma_l.pstex_t} \end{array}} - \ensuremath{\begin{array}{c}\input{pstex/Padlock-Sigma_l.pstex_t} \end{array}}
	}{\{f\}}
	\]
\caption{The diagrammatic form of the
flow equation for vertices
of the Wilsonian effective action.}
\label{fig:Flow}
\ecf

The lobe on the \lhs\ is the Wilsonian effective action vertex corresponding to the fields, $\{f\}$. On the \rhs\ of the flow equation, we identify $X \DummyKernel Y \equiv \dd^{XY}$, where both $X$ and $Y$ can be either $\ptlsf$ or $\ptlasf$. Since the kernels are always internal lines, we
sum over all realizations of $X$ and $Y$ and integrate over the associated fermionic coordinates.
The kernels attach to vertex coefficient functions which can, in principle, have any number of
additional legs. The rule for determining how many legs each of these vertices has---equivalently, the rule for decorating the diagrams on the \rhs---is that the $n_f$ available legs are distributed in all
possible, independent ways. For much greater detail on the diagrammatics, see~\cite{scalar2,primer,thesis}.

In view of their suggestive structure, the two diagrams on the \rhs\ of the flow equation are often called the classical and quantum terms, respectively. However, it should be noted that whilst the classical term does look like a tree diagram, the vertices have really absorbed quantum fluctuations from the bare scale all the way down to the effective scale.

For what follows, it will be useful to consider the effect of the quantum term, in the massless
case. Since the massless effective propagator ties together a $\ptlsf$ and a $\ptlasf$,
only the \Kahler\ potential survives being operated on by the quantum term. Now, bearing
in mind the representation~\eq{FTK}, suppose that it is $K'^{(n,m)}_1$ and $K^{(n,m)}_1$ 
that are tied together by the kernel, which we take to carry momentum, $k$. There
is now a straightforward argument that we can take $K'^{(n,m)}_1$ and $K^{(n,m)}_1$ 
to go as $D^2$ and $\Dbar^2$, up to some function of $k$. The point is that, when two legs
are tied together by an internal line, we can integrate by parts in superspace. This means
that $K'^{(n,m)}_1$ and $K^{(n,m)}_1$ combine to produce
\[
	D^2 \cdots \Dbar^2,
\]
where the ellipsis is some string of superderivatives. Now, if this string comprises just
$D^2$s or $\Dbar^2$s, then our assertion is immediately verified, on account of~\eq{DDD}.
Suppose instead that the string contains superderivatives with loose spinor indices, which might be contracted \emph{elsewhere} in the diagram (this option was not available in the two-point case discussed earlier). On account of the relationships
\be
	\Dbar_{\alphadot} D_\alpha \Dbar^2 \sim \partial_{\alpha\alphadot}\Dbar^2,\qquad
	D_\alpha D_\beta \Dbar^2 \sim \epsilon_{\alpha\beta} D^2 \Dbar^2, \qquad
	\Dbar^2 D_{\alpha} \Dbar^2 = 0,
\label{eq:useful_strings}
\ee
it is clear that our assertion is true, in complete generality.

\section{The Nonrenormalization Theorem}
\label{sec:NRT}

\subsection{Projectors}

To prove the nonrenormalization theorem, we will construct a projector which, when
acting on the Wilsonian effective action, picks out just the superpotential:
\bea
\label{eq:Proj-f}
\lefteqn{
	\SPProj(y) G(\ptlasf,\ptlsf) \equiv
}\\
\nonumber
&&
	\left[
		1
		- y \FourInt{\rho_1} \fder{}{\ptlsf(0,\rho_1,\rhobar_1)}
		+ \frac{y^2}{2!} 
		\FourInt{\rho_1} \FourVol{\rho_2} (\rho_2 \rho_2) 
		\fder{}{\ptlsf(0,\rho_1,\rhobar_1)}
		\fder{}{\ptlsf(0,\rho_2,\rhobar_2)}
		-\cdots
	\right] \left.G\right|_{\ptlsf,\ptlasf= 0}.
\eea
This projector is inspired by Hasenfratz \& Hasenfratz~\cite{H+H} who constructed a
similar projector in scalar field theory, with a view to projecting out the local
potential.

To see how this works, let us first consider its action on the superpotential,
as given by~\eq{FTSP}. To this end, we note from~\eq{FT_D} and~\eq{Ferm-deltafn} that
\be
	(\rho \rho) \Dbar^2(0,\omega,\anti{\omega},\rho,\rhobar) 
	= - \delta^{(4)}(\omega) \delta^{(4)}(\rho).
\label{eq:D^20}	
\ee
Therefore,
\begin{align}
\nonumber
	\SPProj(y) f[\ptlsf] & =
	+4 \sum_{n=2}^{\infty} \frac{4^{n-2} f^{(n)} y^n}{n!}
	\left[
		\prod_{j=1}^n \FourVol{\rho_j} 
	\right]
	\delta^{(4)}(\rho_2) \cdots \delta^{(4)}(\rho_n) \hat{\delta}(0)
	\FourInt{\omega_{23}}
	\cdots
	\FourVol{\omega_{n-1\,n}}
\\ &
\nonumber
	\qquad
	\delta^{(4)}(\rho_1-\omega_{23}) 
	\delta^{(4)}(\omega_{23}-\omega_{34}) \cdots 
	\delta^{(4)}(\omega_{n-2\,n-1} - \omega_{n-1\,n})
	\delta^{(4)}(\omega_{n-1\,n})
\\
	&= 4 \sum_{n=2}^{\infty} \frac{4^{n-2} f^{(n)} y^n}{n!} \hat{\delta}(0)
	\equiv - f(y)\hat{\delta}(0),
\end{align}
where the ill-defined $\hat{\delta}(0)$ can always be regularized at intermediate stages
by working in a finite-sized box.

In~\eq{Proj-f}, it is crucial that the number of  $(\rho\rho)$ factors is one less
than the number of functional derivatives. Had we included an extra such factor in each
term, the projector would have yielded zero. Let us now analyse the
effect of the projector on the \Kahler\ potential, noting
that each $K^{(n,m)}_j$ possesses some
combination of superderivatives, in addition to the necessary $\Dbar^2$, arranged in some
order:
\be
	K^{(n,m)}_j(0,\omega,\anti{\omega},\rho_j,\rhobar_j)
	= -4 
	\delta^{(2)}(\anti{\omega} )  \delta^{(2)}(\rhobar_j )
	\TwoInt{\theta} e^{-i(\omega \theta)} \cdots e^{-i(\rho_j \theta)},
\label{eq:K0}
\ee
where the ellipsis represents some combination of superderivatives (including overall constants)---beyond the $\Dbar^2$ which is always present, and whose effects have been taken into account. Note that all superderivatives are evaluated
at zero momentum, since we have set the first argument of $K_j^{(n,m)}$ equal to zero. Therefore,
\be
	\fermprod{\rho_j}{\rho_j} K^{(n,m)}_j(0,\omega,\anti{\omega},\rho_j,\rhobar_j) 
	\propto \delta^{(4)}(\omega) \delta^{(4)}(\rho_j),
\label{eq:K0rr}
\ee
as we now explain in more detail. First, observe that the $\fermprod{\rho_j}{\rho_j}$ 
converts the $\delta^{(2)}(\rhobar_j )$ of~\eq{K0} into $\delta^{(4)}(\rhobar)$. Consequently,
the $e^{-i(\rho_j \theta)}$ can be replaced with unity. Now, the only combination of superderivatives
evaluated at zero momenta acting on unity which yields a non-zero answer is the trivial combination of no superderivatives. This means that the $\theta$-integral can be performed, yielding
the \rhs\ of~\eq{K0rr}.
Similarly,
\be
	\fermprod{\rhobar'_j}{\rhobar'_j} K'^{(n,m)}_j(0,\omega,\anti{\omega},\rho'_j,\rhobar'_j) 
	\propto \delta^{(4)}(\omega) \delta^{(4)}(\rho'_j).
\label{eq:K0rr-b}
\ee

Thus we find that:
\be
	\SPProj(y) K[\ptlasf,\ptlsf] \propto  \sum_{m=2}^\infty \frac{ y^m}{m!}
	\FourInt{\rho_1} \FourVol{\omega_{12}} 
	K^{(0,m)}_1(0,\omega_{12}, \anti{\omega}_{12},\rho_1,\rhobar_1)
	\delta^{(4)}(\omega_{12}) \hat{\delta}(0) = 0,
\label{eq:K-sf-only}
\ee
as follows from~\eq{K0}.
Consequently, acting on the entire action, our projector does indeed pick out just
the superpotential.

Before moving on, it is worth noting
that a particularly effective and powerful approximation
scheme within the ERG is the derivative expansion (see~\cite{B+B} for a review of the literature,
and~\cite{TRM-Elements} for the key ideas), 
whereby the action is expanded in powers of derivatives. With this in mind, it
is tempting to mimic this in the supersymmetric case and thus construct
a `superderivative expansion'. 
We write the \Kahler\ potential as
\be
	K_\Lambda [\asuf,\suf] \sim \FourInt{x}  \FourVol{\theta} V_\Lambda(\asuf,\suf) +\ldots,
\ee
where $V_\Lambda(\asuf,\suf)$ depends on $\asuf,  \suf$, but not superderivatives thereof, and
the ellipsis indicates terms with extra superderivatives.

We can pick $V$ out of the full \Kahler\ potential by using the projector
\be
	\KPProj(\ybar,y) \equiv \SPProj(y) \anti{\SPProj(y)}
\label{eq:K0-Proj}
\ee
where, of course, we set $\ptlasf, \ \ptlsf  = 0$ after the derivatives from both operators have acted.

Now, a serious health warning should be given. Suppose that we are interested in
searching for pure \Kahler\ fixed points using the superderivative expansion. 
Unfortunately, if we work to lowest order then,
as can be straightforwardly checked, the fixed point equation for $V$ is in fact linear
and, as a consequence, leaves the anomalous dimension entirely undetermined.
Moreover, the reparemtrization invariance of the flow equation is catastrophically
broken. Indeed, as recognized by Wegner~\cite{WegnerInv} and very nicely put by
Morris~\cite{TRM-Elements}, the ERG equation at a fixed point can be thought of
as a non-linear eigenvalue equation for the anomalous dimension. So, the lowest
order in the superderivative expansion looks to be useless for finding fixed points.
Of course, we can always go to higher orders by appropriately generalizing~\eq{K0-Proj}
and, indeed, the resulting coupled equations do become non-linear. Nevertheless,
reparametrization invariance is still broken, and so a unique determination of the
anomalous dimension at a putative non-trivial fixed point is not possible within this
approach. However, this is not something new for Polchinski-style flow equations~\cite{Comellas}
and so it might be profitable to develop this idea further.\footnote{It is interesting to
note that, in scalar field theory, reparametrization invariance can be maintained within
the derivative expansion by using the 1PI flow equation, with a particular form of
cutoff~\cite{TRM-Deriv}. However, there is a price to pay: with this
choice of cutoff function, the derivative expansion does not converge~\cite{TRM-Convergence}!}

\subsection{Proof of the Nonrenormalization Theorem}

We will now prove the nonrenormalization theorem for the massless theory
(the massive case can be done in exactly the same way). To this end, we
apply the projector, $\SPProj(y)$, term by term to the flow equation~\eq{Rescaled}.
The effect on the \lhs\ is obvious. On the \rhs, the most awkward term to deal
with is the quantum one, so we treat this first. However, there are a number of
simplifications we can make. First, it does not make any difference to the following analysis
whether we
take the Wilsonian effective action or seed action contribution to $\Sigma$, so we
just take the former. Secondly, since we are dealing with the massless theory, only
the \Kahler\ potential yields surviving contributions to the quantum term. 
Finally, since we are projecting using $\SPProj(y)$,
the only surviving contributions are those where all external fields are $\ptlsf$. 
Consequently, we wind up with
contributions from the vertices $K^{(1,m)}$ which we split according to~\eq{FTK}:
\begin{multline}
	\SPProj(y) \fder{}{\ptlasf} \cdot c' \cdot \fder{K}{\ptlsf}
	\propto
	\sum_m \frac{y^{m-1}}{(m-1)!}
	\MomInt{4}{k} c'(k^2)
	\FermInt{4}{\kappa} \FourVol{\rho} \FourVol{\omega} \FourVol{\zeta}
	\\
	K'^{(1,m)}_1(-k,\omega,\omegabar, \kappa, \kappabar)
	K^{(1,m)}_1(0,\zeta-\omega,\zetabar- \omegabar, \rho, \rhobar)
	K^{(1,m)}_2(k, \zeta,\zetabar, \kappa, \kappabar).
\label{eq:NR-Pre}
\end{multline}
But we know from the discussion around~\eq{useful_strings} that, since the $K'^{(1,m)}_1$ and
the $K^{(1,m)}_2$ are tied together by a loop integral, we can take them to go
as a $D^2$ and a $\Dbar^2$, respectively, up to some function of $k$. Thus, 
using~\eq{KCTP-mom2b}
we have that
\[
	\FermInt{4}{\kappa}
	K'^{(1,m)}_1(-k,\omega,\omegabar, \kappa, \kappabar)
	K^{(1,m)}_2(k, \zeta,\zetabar, \kappa, \kappabar)
	\propto
	K_0^{\ptlasf\ptlsf}(k,\omega,\omegabar;-k,\zeta,\zetabar).
\]
Furthermore, we have from~\eq{K0} that
\[
	\FermInt{4}{\rho_2} K^{(1,m)}_1(0, \zeta-\omega,\zetabar- \omegabar, \rho_2, \rhobar_2)
	\propto
	A \delta^{(4)}(\zeta-\omega) + B \delta^{(2)}(\zetabar- \omegabar),
\]
for some $A$ and $B$. Therefore, the fermionic integrals in~\eq{NR-Pre} produce
\[
	\FermInt{4}{\omega} \FourVol{\zeta}
	\left[
		A \delta^{(4)}(\zeta-\omega) + B \delta^{(2)}(\zetabar- \omegabar)
	\right]
	K_0^{\ptlasf\ptlsf}(k,\omega,\omegabar;-k,\zeta,\zetabar) = 0,
\]
as can be easily checked by using~\eq{CTP-asf-sf-Explicit}.

The classical terms are easy to project on to with $\SPProj(y)$. First we note that,
because we are working in the massless case, the effective propagator must link
a $\ptlsf$ to a $\ptlasf$, and so at least one of the vertices must be \Kahler\ in order
to end up with a contribution possessing external fields of all one type.
It is simple to check that the classical terms do not yield any contributions
to the superpotential, and so the nonrenormalization theorem is satisfied.

Note that, at a heuristic level, we can see that the nonrenormalization theorem
must be true, just by counting superderivatives.
Ignoring one-point vertices, for the moment, every vertex must possess at least one $D^2$ or $\Dbar^2$. Furthermore, every $n$-point vertex must have a combined number of $D^2$s and $\Dbar^2$s which is at least $n-1$. Now, diagrams generated by the classical term of the flow equation have two vertices with, say, $n$ and $m$ legs, each of which has had one field differentiated. Therefore, the diagram has a combined number of at least $n+m-2$ $D^2$s and $\Dbar^2$s, which is at least equal to the number of external fields. To stand any chance of generating contributions to the superpotential, we must remove enough of the $D^2$s and $\Dbar^2$s,
such that the remaining combined number is $n+m-3$, without ending up with any positive powers of momenta. The only way to perform this removal is via the relationship~\eq{DDD}, but this generates two powers of momentum. Quasi-locality of the vertices means that this cannot be cancelled by negative powers of momenta in the vertices. Since the flow equation involves the differentiated effective propagators, rather than the effective propagators themselves, no negative powers of momenta appear on the internal lines.
Consequently, the classical term in the flow equation cannot generate contributions to the superpotential. 

The diagrams generated by the quantum term in the flow equation have $n$ legs and a combined number of at least $n+1$ $D^2$s and $\Dbar^2$s. Again, we see that it is impossible to generate contributions to the superpotential. 

Were we to include one-point vertices, the discussion for the quantum term remains the same, since vertices contributing to such diagrams must have at least two legs (corresponding to the two ends of the ERG kernel). As for the classical term, diagrams involving a one-point vertex vanish. A one-point vertex  carries zero momentum and, since it necessarily belongs to the superpotential, carries a $\delta$-function in its external fermionic coordinates. Clearly, two one-point vertices yield zero upon mutual attachment. If a one-point vertex attaches to any other vertex, then we can always integrate by parts in superspace to ensure that a $D^2$ or $\Dbar^2$ is explicitly associated with the attachment [in the massive case, these superderivatives could also occur as part of the internal line, as in~\eq{FE-psf-psf}]. 
From~\eqs{D2-Op}{Dbar2-Op}---and remembering to set the momentum to zero---it is clear that such an attachment yields zero.

Since we have rescaled the fields, a flow of the superpotential is induced.
For the flow equation~\eq{Flow_2}, where we recall that we have rescaled using
first $Z$ and then $\lambda(\Lambda)$, the  
classical action comes with an overall $1/\lambda^2$. Using the flow 
 equation, together with the nonrenormalization theorem, it follows that
\be
	\tilde{\gamma} = - \frac{4\beta}{3\lambda},
\label{eq:beta-gamma}
\ee
where the $\beta$-function is defined according to~\eq{betadefn}.

Alternatively, using the flow equation~\eq{Flow} or~\eq{Rescaled}, 
where the rescaling by $\lambda$ is not
performed, we find the more familiar relationship
\be
	\gamma = \frac{2 \beta}{3 \lambda}.
\ee

\section{The Dual Action}
\label{sec:dual}

In~\cite{Trivial}, the key object in the demonstration of the triviality of scalar field theory in $d \geq 4$ is the `dual action', the construction and properties of which we now recall. Denoting the scalar field by $\varphi$, and the effective propagator by $\Delta$, the dual action is defined according to
\[
	- \dual[\varphi] = \ln
	\left\{
		\exp
		\left(
			\hf \fder{}{\varphi} \cdot \Delta \cdot \fder{}{\varphi}
		\right) 
		e^{-\SR[\varphi]}
	\right\}.
\]
In the case that the flow equation is \emph{strictly} the Polchinski equation, there is a simple relationship
between the dual action vertices, $\dualv{n}$, and the correlation functions, $G$:
\begin{align*}
	G(p_1,\ldots,p_n)
	& =
	- \dualv{n}(p_1,\ldots,p_n) \prod_{i=1}^{n} \frac{1}{p_i^2}, \qquad n>2,
\\
	G(p) & = \frac{1}{p^2} \left[1 -  \dualv{2}(p) \frac{1}{p^2} \right].
\end{align*}
However, for other flow equations, these relationships no longer hold. In \sec{rescale}, we introduced a supersymmetric ERG which has a convenient form after scaling the field strength renormalization out of the field. Let us consider the analogue of this equation in the non-supersymmetric case. Now it turns out that the above relationship between the two-point dual action vertex and the two-point correlation function only holds for small momentum.\footnote{This point was not made in versions one through five of~\cite{Trivial}. It does not affect any of the results, however, on the one hand because this is only really important for the interpretation of the dual action and on the other because the various analyses are anyway performed at small momenta.} Moreover, if we were to take a non-trivial seed action, then the simple relationships between the dual action vertices and the correlation functions are greatly complicated. Mindful of these points, we stick to the terminology `dual action', rather than conflating it with the correlation functions.

Returning to the supersymmetric case,  the dual action is defined according to
\be
	- \dual_m[\ptlasf,\ptlsf] = \ln
	\left\{
		e^{\mathcal{Y}_m[\delta/\delta \ptlasf, \delta/\delta \ptlsf] }
		e^{-\SR[\ptlasf,\ptlsf]}
	\right\},
\label{eq:dual}
\ee
where
\be
	\mathcal{Y}_m[\delta/\delta \ptlasf, \delta/\delta \ptlsf] \equiv 
	\fder{}{\ptlasf} \cdot \Delta^{\ptlasf \ptlsf} \cdot \fder{}{\ptlsf}
	+
	\hf \fder{}{\ptlasf} \cdot \Delta^{\ptlasf \ptlasf} \cdot \Dbar^2 \cdot \fder{}{\ptlasf}
	+
	\hf \fder{}{\ptlsf} \cdot \Delta^{\ptlsf \ptlsf} \cdot D^2 \cdot \fder{}{\ptlsf}
\label{eq:construction}
\ee
and the subscript $m$ reminds us that we are working with the massive theory, implying
the presence of the second and third terms on the \rhs. In the massless case, we
define
\[
	\dual[\ptlasf,\ptlsf] = \lim_{m_0 \rightarrow 0} \dual_m[\ptlasf,\ptlsf].
\]

The construction~\eq{construction} has in mind the flow equation~\eq{Flow}, and so the fields in~\eq{dual}
have been rescaled.
Note that, if we also rescale the superspace
coordinates to arrive at flow equation~\eq{Rescaled} (or its massive counterpart),
then the form of the dual action stays the same. However, if we work with
the flow equation~\eq{Flow_2}, we must introduce
\be
	- \dual_{m,\lambda}[\ptlasf,\ptlsf] = \ln
	\left\{
		e^{\mathcal{Y}_{m,\lambda}[\delta/\delta \ptlasf, \delta/\delta \ptlsf] }
		e^{-\SR[\ptlasf,\ptlsf]}
	\right\},
\label{eq:dual-lambda}
\ee
with
\be
	\mathcal{Y}_{m,\lambda}[\delta/\delta \ptlasf, \delta/\delta \ptlsf] 
	\equiv
	\mathcal{Y}_m[\lambda \delta/\delta \ptlasf, \lambda \delta/\delta \ptlsf].
\ee

Let us now compute the
flow of the dual action, using~\eq{Flow}:
\begin{multline}
	-
	\left[
		\flow 
		+ 
		\frac{\gamma}{2} \left(\ptlsf \cdot \fder{S}{\ptlsf} + \ptlasf \cdot \fder{S}{\ptlasf} \right)
	\right]
	\dual_m =
\\
	\gamma
	\left(
		\ptlasf \cdot D^2 \cdot c^{-1} \cdot \Dbar^2 \cdot \ptlsf
		+2m_0 \ptlsf \cdot c^{-1} \cdot \Dbar^2 \cdot \ptlsf
		+2m_0 \ptlasf \cdot c^{-1} \cdot D^2 \cdot \ptlasf
	\right)
\\
	+
	\left[
	e^{\dual_m} 
	\left(
		\ptlsf \cdot \Dbar^2 + 4m_0 \ptlasf
	\right)
	\cdot
	e^{\mathcal{Y}_m}
	c^{-1} \cdot D^2 \cdot
	\left(
		\dd^{\ptlasf\ptlsf} \cdot \fder{\hSR}{\ptlsf}
		+\dd^{\ptlasf\ptlasf} \cdot \Dbar^2 \cdot \fder{\hSR}{\ptlasf}
	\right)
	+ \mathrm{OSC}
	\right]\!.
\label{eq:dualflow}
\end{multline}
Notice that the seed action contributions are restricted to just one term (and its conjugate). Although
other seed action terms are generated, they cancel amongst themselves---either directly,
or courtesy of the relationship
\be
	\fder{}{\ptlasf} \cdot \dd^{\ptlasf \ptlsf} \cdot \frac{\Dbar^2 D^2}{16p^2} 
	\cdot \fder{\hSR}{\ptlsf} e^{-\SR}
	=
	- \fder{}{\ptlasf} \cdot \dd^{\ptlasf \ptlsf} \cdot \fder{\hSR}{\ptlsf} e^{-\SR}.
\label{eq:extra_p^2}
\ee
This follows because, in order to give a non-vanishing contribution, $\hSR$ must
possess at least one $\Dbar^2$ (recall that the $1/p^2$ is nullified by the derivative of the cutoff function
in $\dd^{\ptlasf \ptlsf}$ and so if the $\delta /\delta \ptlsf$ strikes a one-point superpotential vertex,
the entire term just vanishes). Integrating by parts in superspace, we can always ensure
that this $\Dbar^2$---with no further superderivatives---is associated with the leg
hit by the functional derivative. Then we use~\eq{DDD}, remembering that the
$\Dbar^2$ left over belongs to the vertex.

As an aside, it is well worth mentioning that, in the past, cancellations
of the seed action were demonstrated
using elaborate (though increasing sophisticated) 
diagrammatics~\cite{scalar2,primer,NonRenorm,mgiuc,qcd,mgierg1,qed}.
However, as recognized in~\cite{Trivial}, by employing the dual action,
these cancellations can
instead be done with a few lines of algebra, as has been done here.

For what follows, we set $\hSR = 0$. As mentioned earlier, we would
ideally like to keep the seed action general, but then it is not known
how to proceed for the calculations we will do. 
If we now introduce the vertices of the dual action, which we
will denote by $\dualvm{i,j}$, then it is clear that the flow for those with $i+j \neq 2$ is
particularly simple (as a direct consequence of taking $\hSR = 0$) and yields:
\be
	\dualvm{i,j}(-p'_1,\ldots,-p'_i,p_1,\ldots,p_j) = Z^{-(i+j)/2} A(-p'_1,\ldots,-p'_i,p_1,\ldots,p_j), 
	\qquad i+j \neq 2
\label{eq:Dual_Vertices}
\ee
where $A$ is independent of $\Lambda$. 

In the following analysis concerning the existence or otherwise of non-trivial fixed points, we will
draw conclusions using the dual action in two ways. On the one hand, we will draw some conclusions
without having to dig around inside the dual action. In fact, we have an example of this already in~\eq{dualflow}. On the other hand, certain conclusions will be drawn by re-expressing the dual action.
To be specific, we will expand the exponential
\[
	e^{\mathcal{Y}_m[\delta/\delta \ptlasf, \delta/\delta \ptlsf]} 
	= \sum_{i=0}^\infty \frac{1}{i!} \left(\mathcal{Y}_m[\delta/\delta \ptlasf, \delta/\delta \ptlsf]\right)^i
\]
and then allow the functional derivatives to act on $e^{-\SR}$ \emph{before} we perform the sum over $i$. 
Interchanging the order of these two operations is potentially dangerous. Part of the (admittedly heuristic) justification for this procedure is that $\SR$ is the full nonperturbative solution to the flow equation. Thus, even though we have performed this interchange of operations, the resulting series certainly contains more than just standard perturbation theory (standard perturbation theory would correspond to \emph{additionally} replacing the full Wilsonian effective action with only its perturbative contributions). Moreover, we will perform a partial resummation
of this series and assume that the resulting series
can, in principle, be (re)summed to give the full nonperturbative answer.
Rather than performing these manipulations directly at the algebraic level, we prefer to use the  diagrammatic representation of the dual action.

From~\eq{dual}, the dual action comprises all connected diagrams built out of
vertices of the interaction part of the Wilsonian effective action and effective propagators
(it is the logarithm which, as usual, ensures connectedness). A selection of terms contributing 
to $\dualv{2}$ [or $\dualvm{2}$], by which we mean all  $\dualv{i,j}$ with $i+j=2$, is shown in \fig{2pt-terms}.
\bcf[h]
	\[
	\dualv{2} = \ensuremath{\begin{array}{c}\begin{picture}(0,0)%
\epsfig{file=pstex/ReducedWEA-2.pstex}%
\end{picture}%
\setlength{\unitlength}{3947sp}%
\begingroup\makeatletter\ifx\SetFigFont\undefined%
\gdef\SetFigFont#1#2#3#4#5{%
  \reset@font\fontsize{#1}{#2pt}%
  \fontfamily{#3}\fontseries{#4}\fontshape{#5}%
  \selectfont}%
\fi\endgroup%
\begin{picture}(358,579)(1629,-672)
\put(1730,-448){\makebox(0,0)[lb]{\smash{{\SetFigFont{11}{13.2}{\rmdefault}{\mddefault}{\updefault}{\color[rgb]{0,0,0}$\SR$}%
}}}}
\end{picture}%
 \end{array}} + \frac{1}{2} \ensuremath{\begin{array}{c}\begin{picture}(0,0)%
\epsfig{file=pstex/Padlock-2.pstex}%
\end{picture}%
\setlength{\unitlength}{3947sp}%
\begingroup\makeatletter\ifx\SetFigFont\undefined%
\gdef\SetFigFont#1#2#3#4#5{%
  \reset@font\fontsize{#1}{#2pt}%
  \fontfamily{#3}\fontseries{#4}\fontshape{#5}%
  \selectfont}%
\fi\endgroup%
\begin{picture}(418,565)(1606,-593)
\put(1727,-457){\makebox(0,0)[lb]{\smash{{\SetFigFont{11}{13.2}{\rmdefault}{\mddefault}{\updefault}{\color[rgb]{0,0,0}$\SR$}%
}}}}
\end{picture}%
 \end{array}} - \ensuremath{\begin{array}{c}\input{pstex/Dumbbell-2.pstex_t} \end{array}} 
	-\frac{1}{2} \ensuremath{\begin{array}{c}\input{pstex/TP-TL.pstex_t} \end{array}} + \cdots
	\]
\caption{The first few terms that contribute to $\dualv{2}$. Momentum arguments have been suppressed. Each of the lobes represents a vertex of the interaction part of the Wilsonian effective action.
}
\label{fig:2pt-terms}
\ecf

\section{Critical Fixed Points}
\label{sec:CFPs}

As a first application of the dual action formalism, we will investigate the existence of critical
fixed points. This analysis mimics that of~\cite{Trivial}, but with a few small modifications.
To this end, we set the mass to zero (as we must, since we are looking for critical fixed points) and work with dimensionless variables.
Thus our flow equation for the dual action becomes:
\be
	\left[
		\partial_t 
		- \frac{\gamma}{2} 
		\left(
			\ptlsf \cdot \fder{}{\ptlsf} + \ptlasf \cdot \fder{}{\ptlasf} 
		\right)
		+ \frac{\Delta_D}{2} -2
	\right] \dual[\ptlasf,\ptlsf]
	=
	\gamma \ptlasf \cdot D^2 \cdot c^{-1} \cdot \Dbar^2 \cdot \ptlsf
\label{eq:Rescaled-dual-flow}
\ee
Noting
that in rescaled variables $c = c(p^2)$ is independent of $\Lambda$, it
is apparent from the definition~\eq{dual} that, if $m_0 = 0$, then~\eq{FP} implies
\be
	\partial_t \dual_\star[\ptlasf,\ptlsf] = 0.
\label{eq:FP-Dual_condition}
\ee
Now, let us solve~\eq{Rescaled-dual-flow} for the two-point dual action vertex, 
$\dual^{\ptlasf \ptlsf}_\star$.
To this end, we recall from the end of \sec{WEA} that we can write
\be
	\dual^{\ptlasf \ptlsf}_\star(-p,\rho,\rhobar,p,\kappa,\kappabar) =
	z(p)
	\FermInt{4}{\omega}
	D^2(-p,\omega,\anti{\omega},\rho,\rhobar)
	\Dbar^2(p,\omega,\anti{\omega},\kappa,\anti{\kappa}),
\ee
and so we have:
\be
	\left(-\gamma_\star + 2p^2\der{}{p^2}\right) z(p) = \gamma_\star c^{-1}(p^2).
\ee
This equation has solution
\be
	z(p) =
	p^{2\gamma_\star/2}
	\left[
		\frac{1}{b(\gamma_\star)} - \frac{\gamma_\star}{2} 
		\int dp^2 \frac{c^{-1}(p^2)}{p^{2(1+\gamma_\star/2)}}
	\right] \!, 
\label{eq:z-solution}
\ee
where $1/b(\gamma_\star)$ is the (finite) integration constant and is a functional of the cutoff function. 
In the case where $\gamma_\star \neq 0$,
$b$ is defined by the form of $z(p)$ taken if we perform
the indefinite integral by Taylor expanding the cutoff function. For $\gamma_\star = 0$, we make
a choice such that the leading behaviour in the first case coincides with the behaviour
in the second case, as $\gamma_\star \rightarrow 0$.
Thus, for small
momentum, we have
\be
	z(p) = 
	\left\{
	\begin{array}{ll}
	\ds
	\frac{1}{b} p^{2\gamma_\star/2} - \left(1 + \mbox{subleading}\right), & \gamma_\star \neq 0,
	\\[2ex]
	\ds
	\frac{1}{b} -1 , & \gamma_\star = 0.
	\end{array}
	\right.
\label{eq:Soln_2}
\ee
Note that the subleading terms are cutoff dependent, not just with regards to their
prefactors, but also to their structure. For example, if $\gamma_\star = 2$ and
$c'(0) \neq 0$, then the
subleading piece has a nonpolynomial component $p^2 \ln p^2$, but this is absent
altogether if $c'(0) = 0$. However, the
real point to make here is that, so long as $\gamma_\star <2$, the subleading term in the brackets is always subleading compared to $b p^{2(\gamma_\star/2)}$. We will now exclusively
take $\gamma_\star <2$ since, as we will shortly see, this requirement ensures that we
are considering critical fixed points.

The next step
is to introduce the one-particle irreducible (1PI) contributions to the dual action, which
we denote by $\dopiv{i,j}$. At the two-point level we have that $\dual^{\ptlasf \ptlsf}$ is
built up from $\dopi^{\ptlasf \ptlsf}$ according to the geometric series
\begin{multline}
	\dual^{\ptlasf \ptlsf}(-p,\rho,\rhobar;p,\kappa,\anti{\kappa}) =
	\dopi^{\ptlasf \ptlsf}(-p,\rho,\rhobar;p,\kappa,\anti{\kappa}) 
\\
	-
	\FermInt{4}{\omega} 
	\dopi^{\ptlasf \ptlsf}(-p,\rho,\rhobar;p,\omega,\anti{\omega}) \Delta^{\ptlasf \ptlsf}(p) 
	\dopi^{\ptlasf \ptlsf}(-p,\omega,\anti{\omega};p,\kappa,\anti{\kappa}) 
	+ \cdots
\label{eq:geometric}
\end{multline}

Noting that our aim now is to sum the series~\eq{geometric},
we can \emph{schematically} write~\eq{geometric} as:
\[
	\dual^{\ptlasf \ptlsf} = \frac{\dopi^{\ptlasf \ptlsf}}{1 + \Delta^{\ptlasf \ptlsf} \dopi^{\ptlasf \ptlsf}}.
\]
However, this is no more than a mnemonic for~\eq{geometric}, due to the fermionic integrals that must
be performed. 
To perform these integrals, we recall from the end of \sec{WEA} that an arbitrary two-point vertex
can be written as a single $D^2$ and single $\Dbar^2$, up to powers of momentum.
Applying~\eq{DDD}, we see that we can remove all of these $D^2$s and $\Dbar^2$s, with the exception of those on the lines which are external \wrt\ $\dual^{\ptlasf \ptlsf}$, at the expense of a factor of $-16p^2$ for each $\Delta^{\ptlasf \ptlsf}(p)$. Up to the minus sign, this cancels the $1/16p^2$ coming from each effective propagator, in each case leaving behind a $c(p^2)$.

Denoting what is left after we strip off the external $D^2$ and $\Dbar^2$ from
$\dopi^{\ptlasf \ptlsf}$ by $\ztilde$,
we now really can write
\be
	z(p) = \frac{\ztilde(p)}{1 - c(p^2) \ztilde(p)},
\label{eq:z}
\ee
which can be inverted to yield:
\be
	\ztilde(p) = \frac{z(p)}{1 + c(p^2) z(p)}.
\label{eq:invert}
\ee
The final ingredient that we will need is the dressed effective propagator, defined
according to
\be
	\dep^{\ptlasf \ptlsf}(p) \equiv \frac{\Delta^{\ptlasf \ptlsf}(p)}{1 - c(p^2) \ztilde(p)}.
\label{eq:dep}
\ee
At a fixed point with $\gamma_\star < 2$ we find the following small
momentum behaviour
\be
	\dep^{\ptlasf \ptlsf}_\star(p) \sim \frac{1}{p^{2(1-\gamma_\star/2)}},
\ee
which is exactly what we expect at a critical 
fixed point.

Now, the dressed effective propagator can be used to resum sets of loop
diagrams contributing to $\ztilde$, such that all internal lines become dressed, as indicated in
\fig{dressed-TP}.
\bcf[h]
	\[
	\ztilde = \ensuremath{\begin{array}{c}\begin{picture}(0,0)%
\epsfig{file=pstex/ReducedWEA-2-strip.pstex}%
\end{picture}%
\setlength{\unitlength}{3947sp}%
\begingroup\makeatletter\ifx\SetFigFont\undefined%
\gdef\SetFigFont#1#2#3#4#5{%
  \reset@font\fontsize{#1}{#2pt}%
  \fontfamily{#3}\fontseries{#4}\fontshape{#5}%
  \selectfont}%
\fi\endgroup%
\begin{picture}(358,579)(1629,-672)
\put(1730,-448){\makebox(0,0)[lb]{\smash{{\SetFigFont{11}{13.2}{\rmdefault}{\mddefault}{\updefault}{\color[rgb]{0,0,0}$\SR$}%
}}}}
\end{picture}%
 \end{array}} + \frac{1}{2} \ensuremath{\begin{array}{c}\begin{picture}(0,0)%
\epsfig{file=pstex/Padlock-2-dressed-strip.pstex}%
\end{picture}%
\setlength{\unitlength}{3947sp}%
\begingroup\makeatletter\ifx\SetFigFont\undefined%
\gdef\SetFigFont#1#2#3#4#5{%
  \reset@font\fontsize{#1}{#2pt}%
  \fontfamily{#3}\fontseries{#4}\fontshape{#5}%
  \selectfont}%
\fi\endgroup%
\begin{picture}(439,595)(1590,-608)
\put(1727,-457){\makebox(0,0)[lb]{\smash{{\SetFigFont{11}{13.2}{\rmdefault}{\mddefault}{\updefault}{\color[rgb]{0,0,0}$\SR$}%
}}}}
\end{picture}%
 \end{array}}
	-\frac{1}{2} \ensuremath{\begin{array}{c}\input{pstex/TP-TL-dressed-strip.pstex_t} \end{array}} + \cdots
	\]
\caption{Resummation of diagrams contributing to $\ztilde$: the thick lines represent
dressed effective propagators, \eq{dep}, and the stops at the ends of the external lines
indicate that the external $D^2$ and $\Dbar^2$ have been stripped off from each diagram.}
\label{fig:dressed-TP}
\ecf

\subsection{$\gamma_\star \geq 0$}
\label{sec:>0}

In the case where $\gamma_\star = 0$, we know from Pohlmeyer's theorem that the only
critical fixed point is the Gaussian one, So let us now consider critical fixed points with
$\gamma_\star >0$. 
Immediately, this rules out a superpotential at a putative fixed point, as a consequence of
the nonrenormalization theorem.
Next we note that,
from~\eqs{Soln_2}{invert},
\be
	\ztilde_\star(p) = -b p^{-2\gamma_\star/2} + 1 + \cdots.
\label{eq:zt-small_p_1}
\ee
Secondly, we recognize that, by considering the diagrammatic expression for $\ztilde$,
\be
	\ztilde_\star(p) = \mbox{constant} + f(p),
\label{eq:zt-small_p_2}
\ee
where $\lim_{p \rightarrow 0}  f(p) = 0$.
This follows from power counting,
so long as we assume that the Wilsonian effective action vertices are Taylor expandable for small momenta---this being one of our requirements for physical acceptability.
Given $I$ internal lines
and $V$ vertices, there are $L = I -V +1$ loops. If we temporarily ignore the superderivatives
associated with each of the internal legs of the vertices, then the degree of IR divergence is
\[
	\mathbb{D}' \geq 4(I - V + 1) - 2(1-\gamma_\star/2) I,
\]
where we understand $\mathbb{D}' > 0$ to be IR safe. Now, since all two-point vertices have
been absorbed into the dressed effective propagators, each vertex must have at least three legs.  
Given that there are two external legs, this implies that
\[
	I \geq \frac{3V}{2}-1.
\]
Consequently [for $4 \geq 2(1-\gamma_\star/2)$], we have
\[
	\mathbb{D}' \geq -V + 2 + \left( \frac{3V}{2}-1 \right) \gamma_\star.
\]
However, now we must take account of the internal superderivatives in each diagram.

This is easy to do. Let us denote the corrected degree of divergence by $\mathbb{D}$.
Since we are interested in the smallest possible value of $\mathbb{D}$,
we need only consider diagrams built out of three-point vertices: taking vertices with more legs either leaves $\mathbb{D}$ unchanged,  if pairs of these legs are tied together, or increases it 
if the legs attach to other vertices. Similarly, we can consider the minimal number of superderivatives,
amounting to one pair per leg.
Now, from~\eq{FT_D}, we see that the $i$th leg---either internal or external---in some diagram carries
$6-2P_i$ Grassmann numbers, where $P_i$ is the number of powers
of momentum taken on the given leg. However, from~\eq{FTK}, each vertex---being 
three-point---contains an integral
over a pair of dummy coordinates. Thus, the total number of
Grassmann numbers is
\[
	\sum_{i=1}^{3V} (6 -2P_i) - 8V.
\]
Next we notice that, from~\eq{CTP-asf-sf-Explicit}, 
a diagram with an external $D^2$ and an external $\Dbar^2$,
in which the external momentum has been set to zero [\cf~\eq{zt-small_p_2}],
has 8 external Grassmann numbers. Thus, the total number of internal Grassmann numbers
is
\[
	\sum_{i=1}^{3V} (6 -2P_i) - 8(V+1).
\]
However, since the external momentum is set to zero, we can set $P_{3V-1} $ and $P_{3V}$---these
being the $P_i$ we choose to associate with the external legs---to zero. Thus leaves
\[
	\sum_{i=1}^{3V-2} (6 -2P_i) - 8(V+1) + 12.
\]
Now, each internal line contains a fermionic integral, each one of which counts $-4$ 
Grassmann numbers. Therefore, for the diagram not to vanish,
we must equate
\[
	\sum_{i=1}^{3V-2} (6 -2P_i) -8V + 4 = 4I = 6V-4 
	\qquad
	\Rightarrow
	\qquad
	\sum_{i=1}^{3V-2} (6 -2P_i) = 14V - 8.
\]
Consequently, the total number of powers of internal momenta is
\be
	\mathbb{P} \equiv \sum_{i=1}^{3V-2} P_i = 2(V-1),
\label{eq:internalmom}
\ee
yielding a corrected degree of divergence
\[
	\mathbb{D} \geq V + \left( \frac{3V}{2}-1 \right) \gamma_\star.
\]
Given that we are considering $\gamma_\star > 0$, it is obvious that this is always positive,
and so all of our diagrams are IR safe, confirming~\eq{zt-small_p_2}.

It is therefore apparent that, for $\gamma_\star > 0$, \eqns{zt-small_p_1}{zt-small_p_2} are inconsistent and so we conclude that there are no non-trivial fixed points
with $\gamma_\star > 0$ [note that, from~\eq{Soln_2}, $b=0$ is not acceptable, since this would mean that $z(p)$ is singular]. Pohlmeyer's theorem, of course, rules out non-trivial fixed points with
$\gamma_\star = 0$, meaning that, at this stage of the analysis, 
if any non-trivial fixed points are to exist, then they must
have negative anomalous dimension.

Note that if we were to consider diagrams possessing vertices belonging to the
superpotential, then the degree of divergence is lowered, since
superpotential vertices lack (at least) one $D^2$ or one $\Dbar^2$ compared
to \Kahler\ potential vertices. Although this observation is of no use here, since we know
that there cannot be a superpotential at a critical fixed point with $\gamma_\star >0$ (as discussed in the introduction),
this observation will be of use in \sec{beta} where we will
find that it is precisely those diagrams which are IR divergent that contribute to the
one and two-loop $\beta$-function of the Wess-Zumino model. 

\subsection{$\gamma_\star <0$}

Again, this analysis is based on that in~\cite{Trivial}, but with some minor modifications. The vital property of fixed points with negative anomalous dimension, which we
will now exploit, is that
\be
	\lim_{p\rightarrow 0} \ztilde_\star(p) = 1,
\label{eq:minus_one}
\ee
completely independently of the shape of the cutoff function. Note that
for fixed points with positive anomalous dimension, the \rhs\ of~\eq{minus_one}
instead diverges.

The next step is to further resum the diagrams in \fig{dressed-TP}. We cannot do anything
with the first two diagrams. However, the third can be resummed such that the vertices are replaced
with $\dopiv{4}$s. Actually, as discussed in~\cite{Trivial}, this double counts certain contributions
but, crucially, these diagrams are also built entirely out of $\dopiv{n}$s. Thus, we arrive
at the expression in \fig{2pt-dressedx2}.
\bcf[h]
	\[
	\ztilde = \ensuremath{\begin{array}{c}\begin{picture}(0,0)%
\epsfig{file=pstex/ReducedWEA-2-strip.pstex}%
\end{picture}%
\setlength{\unitlength}{3947sp}%
\begingroup\makeatletter\ifx\SetFigFont\undefined%
\gdef\SetFigFont#1#2#3#4#5{%
  \reset@font\fontsize{#1}{#2pt}%
  \fontfamily{#3}\fontseries{#4}\fontshape{#5}%
  \selectfont}%
\fi\endgroup%
\begin{picture}(358,579)(1629,-672)
\put(1730,-448){\makebox(0,0)[lb]{\smash{{\SetFigFont{11}{13.2}{\rmdefault}{\mddefault}{\updefault}{\color[rgb]{0,0,0}$\SR$}%
}}}}
\end{picture}%
 \end{array}} 
	+ \left(
		\frac{1}{2} \ensuremath{\begin{array}{c}\begin{picture}(0,0)%
\epsfig{file=pstex/Padlock-2-dressed-strip.pstex}%
\end{picture}%
\setlength{\unitlength}{3947sp}%
\begingroup\makeatletter\ifx\SetFigFont\undefined%
\gdef\SetFigFont#1#2#3#4#5{%
  \reset@font\fontsize{#1}{#2pt}%
  \fontfamily{#3}\fontseries{#4}\fontshape{#5}%
  \selectfont}%
\fi\endgroup%
\begin{picture}(439,595)(1590,-608)
\put(1727,-457){\makebox(0,0)[lb]{\smash{{\SetFigFont{11}{13.2}{\rmdefault}{\mddefault}{\updefault}{\color[rgb]{0,0,0}$\SR$}%
}}}}
\end{picture}%
 \end{array}} 
		-\frac{1}{6} \ensuremath{\begin{array}{c}\input{pstex/5pt-3pt-dressed.pstex_t} \end{array}}
		+ \cdots 
	\right)
	-\frac{1}{2} \ensuremath{\begin{array}{c}\input{pstex/TP-TL-dressedx2-strip.pstex_t} \end{array}} + \cdots
	\]
\caption{Further resummation of diagrams contributing to $\ztilde$. The brackets contain
terms in which both fields decorate the same vertex. The
second ellipsis represents diagrams built out of $\dopiv{i+j>2}$ vertices.}
\label{fig:2pt-dressedx2}
\ecf

After we take the limit $p \rightarrow 0$, we will denote the first contribution on the \rhs\ of
\fig{2pt-dressedx2} by $w$,
and the rest by $W$ so that, at a fixed point we have
\be
	1 = w + W_\star.	
\label{eq:indep}
\ee
Note that $w$ is a finite number. If $w < 1$ then, as is apparent 
from~\eq{SUSY-split}, 
the \emph{full} action has a kinetic term of the right sign. In this case, $w$ is a free parameter
corresponding to the normalization of the field, with $w=0$ being canonical
normalization.

Let us now suppose that, at a fixed point, $W_\star > 0$ (the following analysis also works
for $W_\star <0$, though such fixed points are already ruled out by our requirement relating
to unitarity).  From~\eq{indep},
we know that $W_\star$ is independent of the shape of the cutoff function.
Now, the only contribution to the cutoff function which is independent of its shape---\ie\ universal---is
$c(0) = 1$. Heuristically, then, we expect that any surviving contributions to $W_\star$ come
from when the loop momenta are precisely equal to zero. To be more precise about this we
note that, since every contribution to $W_\star$ contains at least one loop integral, we can write
\[
	W_\star = \MomInt{D}{k} c(k^2) F[c](k^2).
\]
Given that $\delta W_\star/\delta c =0$ 
 and noting that $W_\star$ is the finite number $1-w$, it appears that we could take
 $F[c](k^2) = c^{-1}(k^2) g(k^2)$, where $\delta g /\delta c = 0$ and the integral over $g$ gives
 $1-w$. However, since $\delta g /\delta c = 0$, we can always choose $c$ such that $F$ diverges
 arbitrarily strongly in the UV. In~\cite{Trivial} it was shown that such behaviour of $F$ is inconsistent.
 Consequently, taking $F[c](k^2) = c^{-1}(k^2) g(k^2)$ is ruled out and that the only option, besides $W_\star=0$, is 
that $F[c](k^2)$ has net contributions only when both $k$, and also all momenta internal
to $F$, are zero.
It is tempting to say that such contributions must have zero support but this does not follow
immediately, as it is quite possible that individual terms contributing to $\ztilde$ diverge as $p \rightarrow 0$. However, as we will now argue, the resummations we  have 
performed in \fig{2pt-dressedx2}  guarantee that there are no such contributions to $W_\star$.

The contributions to \fig{2pt-dressedx2} which might have support for vanishing loop 
momenta are those containing $\dopiv{n}$ vertices, since some of these terms look like
they might possess IR divergences for $p \rightarrow 0$. Now, to show that this does not
occur, we need the momentum dependencies of the $\dopiv{n}_\star$. Let us begin by
noting that~\eq{Rescaled-dual-flow} gives us some useful information about the
$\dualv{n}_\star$. In particular, a dual action vertex with $i$ $\ptlasf$s and $j$ $\ptlsf$s
has a total number of superderivatives
\be
	r_{ij} = 4 + (i+j)\gamma_\star,
\ee
where we recall that momenta can always be written in terms of superderivatives.
Now, since each $\ptlasf$ or $\ptlsf$ necessarily comes with a pair of superderivatives,
we can define the number of `extra' superderivatives by $s$, where
\be
	s_{ij} = 4 + (i+j)(\gamma_\star - 2).
\label{eq:sij}
\ee
However, we are not interested in $s_{i,j}$, per se, but rather the corresponding 
quantity for the $\dopiv{n}_\star$, which we will denote by $\tilde{s}_{i,j}$. To
go from  $s_{ij}$ to $\tilde{s}_{ij}$, we strip off the leg decorations
from the $\dualv{i+j>2}$ and, to this end, define $\dualvpr{i+j>2}$
via
\be
	\dualv{i+j>2}(p_1,\ldots,p_n) = 
	\frac{\dualvpr{i+j>2}(p_1,\ldots,p_n)}{
		\prod_{k=1}^{i+j}\left[1 - c(p_k^2) \ztilde(p_k)\right]
	},
\label{eq:dpr}
\ee
where we have suppressed the fermionic coordinates.
Notice that $\dualvpr{3} = \dopiv{3}$ but, beyond the three point level, there are additional
contributions. However, one of the contributions to $\dualvpr{i+j>2}$ is always
$\dopiv{i+j>2}$ and so, from~\eqs{sij}{dpr}, it is apparent that
\[
	\tilde{s}_{ij} = s'_{i,j} =  4 - (i+j)(\gamma_\star + 2).
\]
When considering two-point diagrams built out of $\dopiv{i+j>2}$s, we know 
from the discussions at the end of  \secs{WEA}{diags}
that all extra superderivatives can be converted into powers of momenta.
Indeed, each vertex effectively comes with
\[
	\frac{\tilde{s}_{ij}}{2} =  2 - (i+j)(\gamma_\star/2 + 1)
\]
`extra' powers of momenta. Thus, we can think of each vertex as coming with
an extra $-(\gamma_\star/2 + 1)$ powers of momentum per leg, plus an additional
two powers.

Ignoring, for the moment, the `necessary' superderivatives, we know to be present,
let us consider the small momentum behaviour, $\mathcal{R}'$, of a diagram
contributing to $\lim_{p \rightarrow 0} \ztilde(p)$
built out of
$V$ $\dopiv{i+j>2}_\star$ vertices and $I$ dressed effective propagators.
Totting up the dependencies from the loop integrals, the dressed
effective propagators and the vertices, we have:
\[
	\mathcal{R}' = 4(I - V + 1) - I(2-\gamma_\star) - (I+1) (\gamma_\star + 2) + 2V 
	= 2(1-V) -\gamma_\star.
\]

Now, just as we did at the end of 
\sec{>0}, we must correct this, to take account of the $D^2$s and $\Dbar^2$s
associated with each of the internal legs (recall that
the external ones have been stripped off). It is straightforward to check that, once
again, the correction is given by~\eq{internalmom} and so we find that
\[
	\mathcal{R} = -\gamma_\star.
\]
Therefore, the diagrams just analysed do indeed go like $p^{-2\gamma_\star/2}$ and so,
for $\gamma_\star <0$, do indeed vanish for $p\rightarrow 0$.

Consequently, the only contributions to $W_\star$ come from the diagrams enclosed by
the brackets in \fig{2pt-dressedx2}.
These are most certainly IR safe for $p \rightarrow 0$, since
the external momentum never flows around any of the loops and so the diagrams do have zero support for 
vanishing loop momenta. Thus,
there are no fixed points with $W_\star \neq 0$.
Therefore, the only fixed points 
with negative anomalous dimension are those for which $w=1$. But, these fixed points
lack a standard kinetic term and so 
correspond to non-unitary theories, upon continuation to Minkowski space.

\section{The $\beta$-function}
\label{sec:beta}

Having made a statement about the space of all possible theories of a scalar chiral superfield---namely that there are no physically acceptable non-trivial fixed-points, we will now return to more familiar territory. To be specific, we will consider the $\beta$-function for the Wess-Zumino model,
considered as a low energy effective theory. 

As is well known, if one chooses a particular set of renormalization schemes, then the
one and two-loop coefficients come out the same. For other renormalization schemes---particularly
those which involve masses--- one gets different numbers which are no less correct (see~\cite{aprop} for a detailed discussion of this point).
It is thus sensible to refer to the one and two loop $\beta$-function coefficients as
pseudo-universal. In what follows, we will consider the massless Wess-Zumino model
in order that we can make a meaningful comparison between
our one and two loop results and the standard pseudo-universal answers. Encouragingly,
we get the correct answers.

It is important to point out that our calculations of the one and two-loop $\beta$-function
are done with general seed action (indeed, this is a nice example of
a case where we know how to proceed without setting $\hSR=0$). Whilst independence
of these pseudo-universal numbers on the seed action is expected, we actually
find more than this: even nonperturbatively, the $\beta$-function turns out to have
no explicit dependence on the seed action. In some sense, this is quite surprising since
the $\beta$-function, beyond two loops, is not even pseudo-universal.

That we do see this unexpected degree of universality seems to be a feature
of the structure of the ERG equation. Indeed, the equation has basically
the same shape irrespective of whether one is considering scalar field
theory, QED, QCD, or the case currently in question. Indeed, the same
degree of universality has been found in these other 
theories~\cite{RG2005,mgiuc,Resum}.

In order to 
compute the $\beta$-function, we must specify the renormalization conditions.
Now, as a consequence of the nonrenormalization theorem, we know that
$\lambda$ is related to the anomalous dimension and the renormalization condition
for $\gamma$ is just that the kinetic term is canonically normalized:
\be
	K = - \frac{1}{\lambda^2} \ptlasf \cdot D^2  \Dbar^2 \cdot \ptlsf +  \cdots,
\label{eq:RC}
\ee
where the ellipsis denotes  contribution of higher dimension operators to
the \Kahler\ potential.
Note that the renormalization condition implies that the $\ptlasf \cdot D^2  \Dbar^2 \cdot \ptlsf$
contribution to the interaction part of the \Kahler\ potential is zero. This is just the
statement that, by $1/\lambda^2$, we mean precisely the coefficient in front of the
complete $-\ptlasf \cdot D^2  \Dbar^2 \cdot \ptlsf$ part of the action. Furthermore,
the three-point superpotential coupling, $f^{(3)}$, is $1/\lambda^2$.

When evaluating the $\beta$-function perturbatively in a theory which is perturbatively renormalizable, but which may be nonrenormalizable beyond perturbation theory, there is a very useful trick we can
use~\cite{scalar1,scalar2,qed}. Namely, we recognize that, as discussed in the introduction,
the Wess-Zumino model is self-similar \emph{at the perturbative level}. In the current
variables, where the canonical dimensions have not been scaled out, this means
that all dependence on $\Lambda$ can either be deduced by na\"{\i}ve power counting
or occurs through $\lambda(\Lambda)$, equivalently $\gamma(\Lambda)$. We will exploit
this below.

Beyond perturbation theory, self-similarity is destroyed, and we must allow for explicit
occurrences of the bare scale, $\Lambda_0$.
Nevertheless, we can still formulate an equation for the $\beta$-function. However, the
above considerations will, at least in principle, affect its evaluation. Actually, as we
will see, the $\beta$-function is in fact free of nonperturbative power corrections
of the form $\Lambda/\Lambda_0$, just as in the manifestly gauge invariant approach to QED~\cite{Resum}, given the definition of the coupling
implicit in the approach~\cite{Resum}.\footnote{When this analysis was first performed in QED, it was speculated whether
resummability of the $\beta$-function in the Wess-Zumino model 
might imply resummability
of the dual action vertices (though this terminology had not yet been
coined). However, there is no reason to expect this to be true.}

\subsection{The $\beta$-Function from the Dual Action}

To derive an expression for the $\beta$-function, we consider the dual action
appropriate to the case where we have rescaled the field by both $\sqrt{Z}$
and $\lambda$---see~\eq{dual-lambda}. 
For the following analysis, we will no longer take the interaction part of the seed action
to be zero and so, in the massless case, we have:
\be
	\left(
		\totalflow + \frac{4\beta}{\lambda} + \tilde{\gamma}
	\right)z_\lambda(p)
	= 
	\left(
		\frac{2 \beta}{\lambda^3} + \frac{\tilde{\gamma}}{\lambda^2}
	\right)
	c^{-1}(p^2/\Lambda^2) + \mbox{seed action term},
\label{eq:beta-1}
\ee
where $z_\lambda$ is defined as what is left after the external $D^2$ and $\Dbar^2$
have been stripped off $\dual_{\lambda}^{\ptlasf \ptlsf}$.

To compute the $\beta$-function, we must employ the renormalization condition~\eq{RC},
and so we are interested in considering~\eq{beta-1} at $p=0$. Now, at first
sight we might worry about strong IR divergences caused by one-particle reducible (1PR)
diagrams; however, the $1/p^2$s in the offending diagrams are compensated by
factors of $p^2$ arising from use of~\eq{DDD}. We might also worry about weaker, logarithmic
IR divergences occurring in loop integrals. These are most certainly present, but cancel
out, as we will discuss in detail below. At intermediate stages of computation, it is perhaps
best to suppose that, term by term, we are looking at both the $\order{p^0}$ and $\order{p^0}\times
\mathrm{nonpolynomial}$ contributions.
Notice that this restriction kills the seed action term. To see this, consider the
seed action term which, up to factors of $\lambda$, can be
read off from~\eq{dualflow} with $m_0=0$. Now, by~\eq{extra_p^2} it is
apparent that the explicitly written $\Dbar^2 D^2$ can be removed, yielding
a factor of $p^2$. Thus, the seed action term contributes at $\order{p^2}$ and $\order{p^2}\times
\mathrm{nonpolynomial}$ and so can be removed from our considerations.
As claimed earlier, we have demonstrated that 
the $\beta$-function has no explicit dependence on
the seed action (there is, of course, implicit dependence buried in the vertices), which
is true nonperturbatively since we have not yet performed a
perturbative expansion of the vertices. 

Recalling~\eq{z}, we introduce the 1PI contribution $\ztilde_\lambda$, appropriate to the flow equation~\eq{Flow_2}, with
\be
	z_\lambda(p) = \frac{\ztilde_\lambda(p)}{1 - \lambda^2 c(p^2/\Lambda^2) \ztilde_\lambda(p)}.
\ee
Utilizing~\eq{beta-gamma},
it is now straightforward to derive the following expression for the $\beta$-function:
\be
	\frac{2 \beta}{3\lambda^3}  +\order{p^2}
		=  -\frac{\totalflow \ztilde_\lambda(p) }{1 + 2\lambda^2 \ztilde_\lambda(p)}.
\label{eq:beta}
\ee
This can be rewritten in the compact form,
\[
	\totalflow \ln
	\left[
		\lambda
		\left(
			1 + \frac{2}{3} \lambda^2  \ztilde_\lambda(p) 
		\right)
	\right] = \order{p^2},
\]
or in the form convenient for computation,
\be
		\frac{2 \beta}{3\lambda^3}  +\order{p^2}
		=  -\frac{\flow \ztilde_\lambda(p) }{1 + 2\lambda^2\ztilde_\lambda(p) + 
			3\lambda^3/2 \partial_\lambda \ztilde_\lambda(p)},
\label{eq:beta-too}			
\ee
where the partial derivative \wrt\ $\Lambda$ is performed at constant $\lambda$.

\subsection{Perturbative Computations}

\subsubsection{The One-Loop Coefficient}

To perform perturbative calculations, we recall~\eq{Pert-action} 
\[
	S \sim \sum_{i=0}^\infty \lambda^{2(i-1)} S_i
\]
and also employ:
\begin{align}
	\ztilde_\lambda(p) & \sim \sum_{i=0}^{\infty} \lambda^{2(i-1)} \ztilde_{\lambda i}(p),
\label{eq:Pert-I}
\\
	\beta & \sim \sum_{i=1}^\infty \lambda^{2i+1} \beta_i.
\label{eq:Pert-beta}
\end{align}

Noting that the one-loop, two-point vertex $K^{\ptlasf \ptlsf}_1$ does not contribute to the $\beta$-function,
as a consequence of the renormalization condition~\eq{RC}, we have:
\be
	\frac{2 \beta_1}{3} + \order{p^2}= -\frac{1}{2}
	\totalflow
	\left[
		\ensuremath{\begin{array}{c}\input{pstex/Padlock-2_0.pstex_t} \end{array}} - \ensuremath{\begin{array}{c}\input{pstex/Thpt_0-Thpt_0.pstex_t} \end{array}}
	\right],
\label{eq:beta1-expr}
\ee
where the zeros inside the vertices denote contributions to the classical action, $S_0$, and 
we recall that the stops on the ends of the external lines indicate that the external $D^2$ and $\Dbar^2$ have been removed.

Let us consider the second diagram, taking the internal momentum to be $k$. Having already extracted the external $D^2$ and $\Dbar^2$ we suppose for the minute that the vertices do not contribute further powers of momenta. 
Temporarily neglecting the fermionic coordinates and overall factors the diagram goes like
\be
	 \left[ \totalflow \MomInt{4}{k}\frac{c^2(k^2/\Lambda^2)}{k^2 (k-p)^2}\right]_{p^0},
\label{eq:proto}
\ee
where we have explicitly indicated the fact that we wish to take the $\order{p^0}$ component, \emph{after} performing the $\Lambda$-derivative [we have taken the liberty of setting $p=0$ in $c((k-p)^2/\Lambda^2)$]. Henceforth, throughout this section, we will use the shorthand
\[
	c_k \equiv c(k^2/\Lambda^2).
\]

There are several ways to evaluate the expression~\eq{proto}~\cite{bo,mgierg2,qed}. However, the most elegant is to recognize that, because the integral is dimensionless, we have the $\Lambda$-derivative of a dimensionless quantity and so for it to survive there must be some scale, besides $\Lambda$, with which to construct a dimensionless function. First we note that the integral is UV finite, due
due to the presence of the cutoff functions, and so no scale can come from here. Secondly, we note
that, as a consequence of \emph{perturbative} self-similarity, there are no hidden couplings / dimensionful quantities buried in the vertices. Consequently, the only place where we can
generate a scale is in the IR, as a consequence of the IR divergences present \emph{before the $\Lambda$ derivative is taken} as $p\rightarrow 0$. In other words, the surviving contributions to~\eq{proto} are of the form:
\[
	\Lambda  \der{}{\Lambda} \ln p^2/\Lambda^2 + \order{p^2}.
\]
With this point in mind, we immediately see that the first diagram of~\eq{beta1-expr} must vanish: there is no IR scale in this diagram.


Let us now include the fermionic coordinates in our analysis of the second diagram in~\eq{beta1-expr}. We will begin by supposing that both vertices belong to the superpotential. For transparency, let us reinstate the external $D^2$ and $\Dbar^2$. The diagram now translates to
\settoheight{\eqnheight}{$\ds \int$}
\begin{multline}
	\hf \totalflow
	\MomInt{4}{k} \FermInt{4}{\rho_1} \FermInt{4}{\rho_2} 
\\
	\left[
	\frac{c^2_k}{16^2 k^2 (k-p)^2}
	F_0^{(3)}(0,\rho,\rhobar;-k,\rho_1,\anti{\rho_1};k,\rho_2,\anti{\rho_2})
	\overline{F}_0^{(3)}(0,\kappa,\anti{\kappa};k,\rho_2,\anti{\rho_2};
		-k,\rho_1,\anti{\rho_1})
	\right]\!,
\label{eq:beta1-contr}
\end{multline}
where we have used~\eq{F^n}, have set $p=0$ in the vertex coefficient functions, and recall that
subscript zeros refer to classical quantities. Now, by the previous arguments, we cannot take any powers of $k$ from the vertices, if we want the diagram to survive. With this in mind, we note that
\begin{align}
\nonumber
	F^{(3)}(0,\rho,\rhobar;0,\omega_1,\anti{\omega_1};0,\omega_2,\anti{\omega_2})
	& = 4^4 \FermInt{4}{\theta}
	\left[
		D^2(0,\theta,\thetabar) e^{i \rho \cdot \theta}
	\right]
	\left[
		D^2(0,\theta,\thetabar) e^{i \omega_1\cdot \theta}
	\right]
	 e^{i \omega_2\cdot \theta}
\\ 
	&=16 \fermprod{\rho}{\rho} \fermprod{\omega_1}{\omega_1} \fermprod{\omega_2}{\omega_2}
	\fermprod{
		(\rhobar + \anti{\omega_1} + \anti{\omega_2})
	}
		{
		(\rhobar + \anti{\omega_1} + \anti{\omega_2})
	},
\end{align}
where we have used the renormalization condition which implies that $f_0^{(3)} = 1$.
Therefore, \eq{beta1-contr} becomes
\be
	\frac{1}{2} \left[\fermprod{\rho}{\rho} \fermprod{\rhobar}{\rhobar}
	\fermprod{\kappa}{\kappa} \fermprod{\anti{\kappa}}{\anti{\kappa}}\right]
	 \totalflow \MomInt{4}{k}  \frac{c^2_k}{k^2 (k-p)^2},
\label{eq:nearly}
\ee
where the contribution in square brackets turns out to be precisely the $\order{p^0}$ contribution to the external $D^2$ and $\Dbar^2$. At this point we note that, were we to have taken either or both of the three-point vertices from the \Kahler\ potential, then the resulting diagram 
would not contribute to $\beta_1$: having arranged the superderivatives such that there are an external $D^2$ and $\Dbar^2$, the diagram would either be too high an order in $p$, or would be killed by the $\Lambda$-derivative, due to additional powers of internal momenta. Combining~\eq{beta1-expr} and~\eq{nearly} with the fermionic coordinates stripped off yields:
\be
	\frac{2\beta_1}{3} + \order{p^2} = \frac{1}{2} \totalflow \MomInt{4}{k}   
	\frac{c^2_k}{k^2 (k+p)^2}.
\label{eq:beta1}
\ee

All that remains to be done is to compute the integral, which does not involve any fermionic coordinates.
There are several ways to do this. The most efficient involves using dimensional regularization, not as a means of regularizing the integral in the UV, but as a trick for extracting the part which survives differentiation \wrt\ $\Lambda$. We emphasise that using dimensional regularization in this way, and at this stage, is entirely valid, does not spoil our superspace implementation, and works to any number of loops (or even nonperturbatively). The key point is that it is simply a trick for evaluating a finite bosonic quantity. Clearly, given that the trick is known to work, the answer to~\eq{beta1} should not depend on the history of how this equation was obtained. For the details of this elegant method, see~\cite{mgierg2,qed}; see~\cite{scalar2} for an alternative technique formulate directly in $d=4$.
It is reassuring that we get the usual result:
\be
	\beta_1 = \frac{3}{2} \frac{1}{(4\pi)^2}.
\label{eq:beta1-ans}
\ee

\subsubsection{The Two-Loop Coefficient}

At the two-loop level, although there are many diagrams which could, in principle,
contribute to the $\beta$-function, only two give non-vanishing contributions:
\be
	\frac{2\beta_2}{3} + \order{p^2} = 
	\hf
	\totalflow
	\left[
		\ensuremath{\begin{array}{c}\input{pstex/Thpt_0x4.pstex_t} \end{array}}
	\right]
	+
	\hf
	\ensuremath{\begin{array}{c}\input{pstex/Thpt_0-Thpt_0.pstex_t} \end{array}} 
	\totalflow \left[\ensuremath{\begin{array}{c}\input{pstex/Thpt_0-Thpt_0.pstex_t} \end{array}}\right],
\label{eq:beta_2-expr}
\ee
where the second term on the \rhs\ comes from 
the second term in the denominator of~\eq{beta-too}
(the third term in the denominator does not contribute until three loops).

As with $\beta_1$, only vertices belonging to the superpotential produce surviving contributions and these can be cast in the form:
\be
	\frac{2\beta_2}{3} + \order{p^2} 
	=	
	\totalflow \MomInt{4}{k} \MomInt{4}{l}
	\left[
		\frac{c^3_k \, c_{l-k}\, c_l}{k^2 (k-p)^2 (l-k)^2 l^2} 
		- \hf \frac{c^2_k}{k^2 (k-p)^2} \frac{c^2_l}{l^2 (l-p)^2}
	\right]
\label{eq:almost-beta2}
\ee

Notice that a relative sign is introduced between the two terms, as compared with~\eq{beta_2-expr}. This comes about as the result of employing~\eq{DDD} along the internal lines carrying the outer loop momentum [taking the outer loop momentum of the first diagram to be $k$, this also explains why the first term in~\eq{almost-beta2} $\sim 1/k^2$, rather than $1/k^4$]. The relative factor of $1/2$ between the two terms, as compared with~\eq{beta_2-expr}, arises from recognizing that both contributions to the second term can be taken inside the derivative, at the expense of a factor of $1/2$. An evaluation of the integrals is given, directly in $d=4$ in~\cite{scalar2}. For details of the alternative method employing dimensional regularization, see~\cite{mgierg2}. Either way, the expected answer is obtained:
\be
\label{eq:beta2-final}
	\beta_2 = -\frac{3}{2}\frac{1}{(4\pi)^4}.
\ee

\subsection{Nonperturbative Considerations}
\label{sec:NPC}

We will now argue, along the lines of~\cite{Resum}, that even in the case where there
is an additional physical scale present, violating self-similarity, the $\beta$-function does not receive
nonperturbative corrections. First of all, let us recall from~\eq{PowerCorr} 
that we can re-express any such terms
using $\lambda$, according to 
\[
	\frac{\Lambda}{\Lambda_0} \sim e^{-1/ 2 \beta_1 \lambda^2(\Lambda)} + \ldots,
\]
where the prefactor contains the $\Lambda_0$ dependence.

Let us now return to the expression for  the $\beta$-function,
\eq{beta-too}, before any perturbative expansion has been performed. Quite irrespective
of whether we now perform a perturbative expansion and whether there are additional
scales floating around, it is still the case that there are nonpolynomial contributions to $\ztilde$
which blow up as $p \rightarrow 0$. Moreover, since the \lhs\ of~\eq{beta-too} is safe in the
$p \rightarrow 0$ limit, it is apparent that  any such divergences must cancel between numerator
and denominator on the \rhs. Therefore, it must
be that we can write:
\be
	\frac{2 \beta}{3 \lambda^3} + \order{p^2} = 
	\frac{F_1(\lambda^2) G(\lambda^2, \ln p^2/\Lambda^2)}{F_2(\lambda^2) G(\lambda^2, \ln p^2/\Lambda^2)} 
	= \frac{F_1(\lambda^2)}{F_2(\lambda^2)},
\label{eq:beta-form}
\ee
where $F_1$, $F_2$ and $G$ are unknown functions.

To begin with, let us reconsider perturbation theory.
Let us suppose that, 
at order $\lambda^{2i}$, the strongest IR divergence carried by $\ztilde(p)$, at
$\order{p^0 \times \mathrm{nonpolynomial}}$, goes like
\be
	\lambda^{2i} \ln^j p^2 /\Lambda^2.
\label{eq:div-pert}
\ee
In the numerator of~\eq{beta-too}, the $\Lambda$-derivative (which we recall is performed at constant $\lambda$) reduces this divergence to one of the form
\be
	\lambda^{2i} \ln^{j-1} p^2 /\Lambda^2
\label{eq:num-div}
\ee
whereas, in the denominator, a contribution of the form
\be
	\lambda^{2(i+1)} \ln^m p^2 /\Lambda^2
\label{eq:den-div}
\ee
is produced.
At first sight, we have found that terms of the form~\eq{div-pert} provide a divergent contribution to the denominator which
does not seem to exist in the numerator. Of course, there is no real problem here: all we need to do is consider diagrams with an extra loop. In such diagrams there are
contributions of the form~\eq{div-pert} but with $i \rightarrow i+1$ and $j \rightarrow j+1$. Terms like this in the numerator are, after differentiation \wrt\ $\Lambda$, of precisely the right form to cancel denominator contributions of the type~\eq{den-div}. 

But now consider a contribution of the type
\be
	\lambda^{2i} e^{-a/\lambda^2} \ln^j p^2 /\Lambda^2,
\label{eq:div-nonpert}
\ee
where again we assume that, for our choice of $i$, there is no stronger IR divergence. In the numerator of~\eq{beta-too} this contributes terms of the form
\be
	\lambda^{2i} e^{-a/\lambda^2} \ln^{j-1} p^2 /\Lambda^2
\label{eq:num-np-div}
\ee
and in the denominator it yields terms of the form
\be
	\lambda^{2i} e^{-a/\lambda^2} \ln^j p^2 /\Lambda^2 + \ldots,
\label{eq:den-np-div}
\ee
where the ellipsis denotes terms higher order in $\lambda^2$. 
(The explicitly written term comes from the last piece of the denominator.)
Crucially, 
\eqs{num-np-div}{den-np-div} are \emph{the same order} in $\lambda^2$.
Since, by assumption, there are no terms in $\ztilde(p)$ which
are of order $\lambda^{2i}e^{-a/\lambda^2}$ but which have a stronger 
IR divergence than~\eq{div-nonpert},
there is no way that the denominator contribution~\eq{den-np-div}
can ever be cancelled. From~\eq{beta-form}, we therefore conclude
that
terms of the
type~\eq{div-nonpert} must be absent from~\eq{beta-form}, unless $j=0$.
But it is easy to see that $j=0$ terms can appear only in $G(\lambda^2,\ln p^2/\Lambda^2)$
and not in $F_1(\lambda^2)$ or $F_2(\lambda^2)$: for if this condition were violated, then
we would necessarily produce contributions of the form~\eq{div-nonpert}, when
we expand out $F_1(\lambda^2) G(\lambda^2, \ln p^2/\Lambda^2)$. In conclusion, 
the only contributions to the $\beta$-function 
of the form~\eq{div-nonpert} that are allowed---namely
those with $j=0$---cancel out!

It is now straightforward to generalize this argument to show that only
the perturbative contributions to the $\beta$-function survive. First, we
note that the above argument is not affected if we consider terms
which include $e^{-b/g^4}$, $e^{-c/g^6}$ \etc, or products of such terms.
Secondly, we can allow additional functions of $g$ to come along
for the ride, so long as they do not spoil the requirement that the ERG
trajectory sinks into the Gaussian fixed point as $\Lambda \rightarrow 0$.

Note that in the massive case  there is no reason to expect the $\beta$-function to be free of 
power corrections, since it is quite consistent to pick up terms like
\[
	\frac{m_0}{\Lambda_0} e^{-a/\lambda^2},
\]
because the mass now regularizes terms which previously diverged as $p \rightarrow 0$.
[Actually, with the presence of more than one type of two-point vertex, even
relationships like~\eq{z} need to be rederived.]
This observation could be important  when inverting the relationship between the
dual action and the Wilsonian effective action:
\be
	-\SR[\ptlasf,\ptlsf] = \ln
	\left\{
		e^{-\mathcal{Y}_m[\delta/\delta \ptlasf, \delta/\delta \ptlsf] }
		e^{-\dual_m[\ptlasf,\ptlsf]}
	\right\}.
\label{eq:recover}
\ee
The point is that, since the dual action vertices are IR divergent, we presumably must take $m_0 \neq 0$,
at least at intermediate stages, in order to make sense of~\eq{recover}. Whilst it is
true that once $\SR$ has been computed, we should be able to safely send $m_0 \rightarrow 0$, 
it is quite conceivable that contributions to the Wilsonian effective action
of the form $m_0/\Lambda_0 \times  \Lambda/m_0$ are generated. Such terms are, of course,
perfectly well defined in the $m_0 \rightarrow 0$ limit.

\section{Conclusion}
\label{sec:conc}

In this paper, we have given a comprehensive treatment of the renormalization of theories of a scalar chiral superfield. Central to our approach is the notion of `theory space' rather than a specific model. Theory space refers to the space of all possible (quasi-local) effective actions and it is in this space that we must search for non-trivial fixed points, the existence of which is necessary if we are to construct interesting theories that are nonperturbatively renormalizable in $d=4$. Unfortunately, just as in the case of scalar field theory, we argued in \sec{CFPs} that there cannot be any non-trivial fixed points satisfying the conditions of quasi-locality and unitarity (upon continuation to Minkowski space). Consequently, it is not possible to write down a physically acceptable non-trivial bare action for which the bare scale can be removed. Thus we conclude that the Wess-Zumino model suffers from the problem of triviality.

Of course, this does not stop one from treating the Wess-Zumino model as an effective theory and we did precisely that in \sec{beta}, where the $\beta$-function was discussed. It is heartening to see that, given familiarity with the formalism, it is no harder to compute the perturbative $\beta$-function within the ERG than within other formalisms. Moreover, the approach has the added benefit of being defined nonperturbatively and we exploited this to show that the $\beta$-function of the massless Wess-Zumino model (corresponding to the natural definition of the coupling within the approach) is in fact free of nonperturbative power corrections.

The other main result of the paper is the proof of the nonrenormalization theorem, directly from the ERG, performed in \sec{NRT}. 

Besides these three results, it should be emphasised that the framework developed in this paper is interesting in its own right. After all, it is formulated directly in $d=4$, has UV regularization built in, is manifestly supersymmetric and is defined nonperturbatively. It is hoped that these features will encourage further research into this subject.

\section*{Note Added}

Since this paper was originally posted, there have been a number of interesting papers applying the ERG to supersymmetric theories~\cite{Synatschke:2008pv,Gies:2009az,Synatschke:2009nm,Synatschke:2009da}.

\appendix

\section{SUSY Conventions}
\label{app:Conventions}

To define the $\mathcal{N} =  1$ superfield formalism in four dimensional Euclidean space, we follow Lukierski and Nowicki~\cite{Lukierski} (see also~\cite{TRM-Euclidean} for a digestible summary). The lowest dimensional faithful spinor representation of $\SO(4)$ is described by two independent $\SU(2)$ spinors, which we will denote
\be
	\theta^{\firstspinor{\alpha}}, \qquad \theta^{\secondspinor{\alpha}}.
\label{eq:so4-spinor}
\ee
Note that, compared to~\cite{Lukierski}, we have taken the indices to be upper, rather than lower, so that our formulae map directly on to those of Wess and Bagger~\cite{W+B}. 
Furthermore, when comparing to~\cite{Lukierski}, the reader should be warned: some of the semicolons of~\cite{Lukierski} are in the wrong place, some are either implicit or actually missing and the odd one has been accidentally replaced by a subscript $j$, which looks remarkably similar. 

The convention for complex conjugation is as follows:
\be
	\left(\theta^{\firstspinor{\alpha}}\right)^{*} = \theta^{\firstspinor{\dot{\alpha}}},
	\qquad
	\left(\theta^{\secondspinor{\alpha}}\right)^{*} = \theta^{\secondspinor{\dot{\alpha}}}.
\label{eq:CC}
\ee
Consequently, the lowest dimension Hermitean Euclidean superspace is
\be
	S = (x_\mu, \theta^{\firstspinor{\alpha}}, \theta^{\secondspinor{\alpha}}, 
			\theta^{\firstspinor{\dot{\alpha}}}, \theta^{\secondspinor{\dot{\alpha}}}),
\label{eq:minimal}
\ee
which corresponds to $\mathcal{N}=2$ supersymmetry~\cite{zumino}. To obtain $\mathcal{N}=1$ superspace, we
restrict ourselves to non-Hermitean `Grassmann-analytic' chiral superspaces:
\be
	S^- = (x_\mu, \theta^{\firstspinor{\alpha}}, \theta^{\secondspinor{\dot{\alpha}}}),
	\qquad
	S^+ = (x_\mu, \theta^{\secondspinor{\alpha}}, \theta^{\firstspinor{\dot{\alpha}}}).
\label{eq:GrassmannAnalytic}
\ee
(The reader should be warned that the labelling of $S^{\pm}$ is not consistent throughout~\cite{Lukierski}.) Although we have lost Hermitean self-conjugacy for $S^+$ and $S^-$, it is replaced by
`Osterwalder and Schrader' (OS) self-conjugacy~\cite{O+S1}, which involves Hermitean conjugation, followed by time ($x_4$) reversal. Under this operation,
\be
	\theta^{\firstspinor{\alpha}} \stackrel{\mathrm{OS}}{\longleftrightarrow} 
		\theta^{\secondspinor{\dot{\alpha}}},
	\qquad
	\theta^{\secondspinor{\alpha}} \stackrel{\mathrm{OS}}{\longleftrightarrow}  
		\theta^{\firstspinor{\dot{\alpha}}}.
\label{eq:OS}
\ee
Euclidean superfields which are OS-conjugate become Hermitean after continuation to Minkowski space and imposition of the Majorana condition. Focussing on $S^+$, the $\sigma$ matrices are chosen such that they are OS self-conjugate:
\be
	\sigma_{\scriptstyle \firstspinor{\dot{\alpha}} \secondspinor{\alpha}}^\mu
	= (\sigma_j, i)_{\firstspinor{\dot{\alpha}} \secondspinor{\alpha}}.
\label{eq:sigma-OS}
\ee
If we now make the following identifications, where a `bar' denotes OS-conjugation:
\be
	\theta^{\firstspinor{\dot{\alpha}}} \equiv \thetabar^{\dot{\alpha}}, \qquad
	\theta^{\secondspinor{\alpha}} \equiv \theta^{\alpha},
	\qquad
	\sigma_{\scriptstyle \firstspinor{\dot{\alpha}} \secondspinor{\alpha}}^\mu
	\equiv \sigma_{\dot{\alpha} \alpha}^\mu,
\ee
then our spinor algebra conventions can be read off from those of Wess and Bagger, so long
as we replace the Minkowski metric with $\delta_{\mu \nu}$ and do not look inside $\sigma^\mu$.

For completeness, we give the various formulae that were used to obtain the results in this paper.
Indices are raised and lowered with the epsilon tensors $\epsilon^{\alpha\beta}$, $\epsilon_{\alpha\beta}$, $\epsilon^{\dot{\alpha} \dot{\beta}}$ and $\epsilon_{\dot{\alpha} \dot{\beta}}$ with $\epsilon_{21} = \epsilon^{12} = 1$ \etc\ Defining
\be
	\sigmabar^{\mu \dot{\alpha} \alpha} \equiv
	\epsilon^{\dot{\alpha} \dot{\beta}} \epsilon^{\alpha \beta} 
	\overline{\sigma}^{\mu}_{\dot{\beta} \beta}
\label{eq:sigmabar}
\ee
we find
\begin{subequations}
\begin{align}
	\left(
		\sigma^\mu \sigmabar^\nu + \sigma^\nu \sigmabar^\mu
	\right)^{\ \beta}_\alpha & = - 2 \delta^{\mu\nu} \delta^{\ \beta}_\alpha,
\\
	\left(
		\sigma^\mu \sigmabar^\nu + \sigma^\nu \sigmabar^\mu
	\right)^{\dot{\alpha}}_{\ \dot{\beta}} & = - 2 \delta^{\mu\nu} \delta^{\dot{\alpha}}_{\ \dot{\beta}},
\end{align}
\end{subequations}
with the completeness relations:
\begin{subequations}
\begin{align}
	\mathrm{Tr} \sigma^\mu \sigmabar^\nu & = - 2 \delta^{\mu \nu}, 
	\\
	\sigma^{\mu}_{\alpha \dot{\alpha}}\sigmabar_{\mu}^{\dot{\beta} \beta} 
	& = -2\delta^{\ \beta}_\alpha \delta^{\ \dot{\beta}}_{\dot{\alpha}}.
\end{align}
\end{subequations}

The spinor summation conventions are:
\begin{subequations}
\begin{align}
\label{eq:upperlower}
	\psi \chi 
	& = \psi^\alpha \chi_\alpha 
	= - \psi_\alpha \chi^\alpha 
	= \chi^\alpha \psi_\alpha = \chi \psi,
	\\
	\anti{\psi} \anti{\chi} 
	& = \anti{\psi}_{\dot{\alpha}} \anti{\chi}^{\dot{\alpha}}
	= -\anti{\psi}^{\dot{\alpha}} \anti{\chi}_{\dot{\alpha}}
	= \anti{\chi}_{\dot{\alpha}} \anti{\psi}^{\dot{\alpha}} 
	= \anti{\chi} \anti{\psi},
\end{align}
\end{subequations}
where we will often enclose spinor products in round brackets, for clarity. We define
\be
	(\rho p \thetabar) \equiv \rho^{\alpha} \sigma^{\mu}_{\alpha\dot{\alpha}} \thetabar^{\dot{\alpha}} p_\mu.
\ee
It should be noted, to avoid possible confusion, that Lukierski et al.\ use what, in our notation, amounts to an `upper-lower' convention of type~\eq{upperlower} for \emph{both} 
$\theta^{\secondspinor{\alpha}}$ and $\theta^{\firstspinor{\dot{\alpha}}}$.
Consequently, whilst our superspace operators $Q$ and $\anti{Q}$, $D$ and $\anti{D}$ take the same form as in Wess and Bagger
\begin{subequations}
\begin{align}
\label{eq:Q}
	Q_\alpha & = \pder{}{\theta^\alpha} - i \sigma_{\alpha \dot{\alpha}}^\mu \thetabar^{\dot{\alpha}} 
			\partial_\mu,
\\
\label{eq:Qbar}
	\anti{Q}_{\dot{\alpha}} & =  -\pder{}{\anti{\theta}^{\dot{\alpha}}} - i \theta^\alpha 
		\sigma_{\alpha \dot{\alpha}}^\mu \partial_\mu,
\\
	D_\alpha & =  \pder{}{\theta^\alpha} + i \sigma_{\alpha \dot{\alpha}}^\mu \thetabar^{\dot{\alpha}} 
			\partial_\mu,
\\
	\anti{D}_{\dot{\alpha}} & =  -\pder{}{\anti{\theta}^{\dot{\alpha}}} - i \theta^\alpha 
		\sigma_{\alpha \dot{\alpha}}^\mu \partial_\mu,
\end{align}
\end{subequations}
they differ from those in~\cite{Lukierski}.

When Fourier transforming the fermionic coordinates, the starting point is to recognize that
\be
	16 \FermInt{4}{\rho} e^{i\rho \cdot (\omega - \theta)} = \delta^{(4)}(\omega - \theta),
\label{eq:FermDeltaFn}
\ee
where
\be
	\delta^{(4)}(\theta) = \fermprod{\theta}{\theta} \fermprod{\thetabar}{\thetabar}
\label{eq:Ferm-deltafn}
\ee
and
\be
	\FermInt{2}{\theta} \theta\theta = 1, \qquad \FermInt{2}{\thetabar} \thetabar\thetabar = 1.
\ee

Some useful formulae are:
\begin{subequations}
\begin{align}
	D^2(-p,\theta,\thetabar) e^{i\rho \cdot \theta}
	& =
	\left[
		-\fermprod{\rho}{\rho} - 2i (\rho p \thetabar) + p^2 \fermprod{\thetabar}{\thetabar}
	\right]
	e^{i\rho \cdot \theta},
\label{eq:D2-Op}
\\
	\Dbar^2(p,\theta,\thetabar) e^{-i\rho \cdot \theta} 
	& =
	\left[
		-\fermprod{\rhobar}{\rhobar} + 2i (\theta p \rhobar) + p^2 \fermprod{\theta}{\theta}
	\right]
	e^{-i\rho \cdot \theta}.
\label{eq:Dbar2-Op}
\end{align}
\end{subequations}

\section{Classical Two-Point Vertices}
\label{app:CTP}

The completely Fourier transformed classical, two-point contribution to the ${\Kah}^{\ptlasf \ptlsf}$ vertex is given by:
\begin{multline}
	\lefteqn{
	{\Kah}_0^{\ptlasf \ptlsf}(-p,\rho,\rhobar;p,\kappa,\anti{\kappa}) = -c^{-1}(p^2/\Lambda^2)
	\left\{
	\left[
		\fermprod{\rho}{\rho} \fermprod{\rhobar}{\rhobar} + 4(\rho p \rhobar) - 4p^2
	\right]
	\left[
		\fermprod{\kappa}{\kappa} \fermprod{\anti{\kappa}}{\anti{\kappa}} + 4(\kappa p \anti{\kappa}) - 4p^2
	\right]
	\right.
	}
	\\
	\left.
	+ 8 (\kappa p \rhobar) \fermprod{\rho}{\rho} \fermprod{\anti{\kappa}}{\anti{\kappa}} 
	+ 16p^2 \fermprod{\rho}{\rho} \fermprod{\rhobar}{\anti{\kappa}} 
	- 16p^2 \fermprod{\rho}{\rho} \fermprod{\anti{\kappa}}{\anti{\kappa}}
	+16p^2 \fermprod{\anti{\kappa}}{\anti{\kappa}}\fermprod{\rho}{\kappa} 
	+ 32p^2 (\rho p \anti{\kappa})
	\right\}\!\!,
\label{eq:CTP-asf-sf-Explicit}
\end{multline}
whilst the classical mass terms are given by:
\begin{subequations}
\bea
\nonumber
	\lefteqn{
		{S}_0^{\ptlsf \ptlsf}(-p,\rho,\rhobar;p,\kappa,\anti{\kappa}) = - 16 m_0 c^{-1}(p^2/\Lambda^2)
	}
	\\
	&&
	\times
	\left\{
	-\frac{1}{4}
	\fermprod{\anti{\kappa}}{\anti{\kappa}} \fermprod{\rhobar}{\rhobar}
	\fermprod{(\rho + \kappa)}{(\rho + \kappa)}
	+\left[p^2 - ((\rho + \kappa) p \anti{\kappa})\right]
	\fermprod{(\rhobar + \anti{\kappa})}{(\rhobar + \anti{\kappa})}
	\right\}
\label{eq:CTP-m-sf}
\\
\nonumber
	\lefteqn{
		{S}_0^{\ptlasf \ptlasf}(+p,\rho,\rhobar;-p,\kappa,\anti{\kappa}) = - 16 m_0 c^{-1}(p^2/\Lambda^2)
	}
\\ 	
	&&
	\times
	\left\{
	-\frac{1}{4}
	\fermprod{\kappa}{\kappa} \fermprod{\rho}{\rho}
	\fermprod{(\rhobar + \anti{\kappa})}{(\rhobar + \anti{\kappa})}
	+\left[p^2 - (\kappa p (\rhobar + \anti{\kappa}))\right]
	\fermprod{(\rho + \kappa)}{(\rho + \kappa)}
	\right\}.
\label{eq:CTP-m-asf}
\eea
\end{subequations}

\bibliography{../../../Biblios/ERG,../../../Biblios/Renormalons,../../../Biblios/SUSY,../../../Biblios/Books,../../../Biblios/Foundations}

\end{document}